\documentclass{emulateapj}
\usepackage{placeins}
\usepackage{graphicx}
\usepackage{amsmath}
\usepackage{mathrsfs}
\shorttitle{Circumplanetary Disk Accretion}
\shortauthors{Zhu et al}

\begin{document}

\title{Shock-driven Accretion in Circumplanetary Disks: Observables and Satellite Formation}

\author{Zhaohuan Zhu\altaffilmark{1}, Wenhua Ju\altaffilmark{2}, James M. Stone\altaffilmark{2}}

\altaffiltext{1}{Department of Physics and Astronomy, University of Nevada, Las Vegas, 
4505 South Maryland Parkway, Las Vegas, NV 89154, USA}
\altaffiltext{2}{Department of Astrophysical Sciences, 4 Ivy Lane, Peyton Hall,
Princeton University, Princeton, NJ 08544, USA}

\email{zhzhu@physics.unlv.edu }

\newcommand\msun{\rm M_{\odot}}
\newcommand\lsun{\rm L_{\odot}}
\newcommand\msunyr{\rm M_{\odot}\,yr^{-1}}
\newcommand\be{\begin{equation}}
\newcommand\en{\end{equation}}
\newcommand\cm{\rm cm}
\newcommand\kms{\rm{\, km \, s^{-1}}}
\newcommand\K{\rm K}
\newcommand\etal{{\rm et al}.\ }
\newcommand\sd{\partial}
\newcommand\mdot{\rm \dot{M}}
\newcommand\rsun{\rm R_{\odot}}

\begin{abstract}
Circumplanetary disks (CPDs) control the growth of planets, supply material for satellites to form, and provide observational signatures of young forming planets. We have carried out two dimensional hydrodynamical simulations with radiative cooling to study CPDs, and suggested a new mechanism to drive the disk accretion. Two spiral shocks are present in CPDs, excited by the central star. We find that spiral shocks can at least contribute to, if not dominate the angular momentum transport and energy dissipation in CPDs. Meanwhile, dissipation and heating by spiral shocks have a positive feedback on shock-driven accretion itself. As the disk is heated up by spiral shocks, the shocks become more open, leading to more efficient angular momentum transport. This shock driven accretion is, on the other hand, unsteady on a timescale of months/years due to production and destruction of vortices in disks. After being averaged over time, a quasi-steady accretion is reached from the planet's Hill radius all the way to the planet surface, and the disk $\alpha$-coefficient characterizing angular momentum transport due to spiral shocks is $\sim$0.001-0.02. The disk surface density ranges from 10 to 1000 g cm$^{-2}$ in our simulations, which is at least 3 orders of magnitude smaller than the ``minimum mass sub-nebula'' model used to study satellite formation;  instead it is more consistent with the ``gas-starved'' satellite formation model. Finally, we calculate the millimeter flux emitted by CPDs at ALMA and EVLA wavelength bands and predict the flux for several recently discovered CPD candidates, which suggests
that ALMA is capable of discovering these accreting CPDs.

\end{abstract}

\keywords{hydrodynamics, accretion, accretion disks, planets and satellites: detection, planets and satellites: formation, planet-disk interactions, shock waves}

\section{Introduction}
 {Planets form and grow in circumstellar disks.} 
Initally when a protoplanet  forms, the solid core is surrounded by a hydrostatic gaseous envelope
that is in contact with the planet's Hill sphere and the rest of the circumstellar disk. 
As the core's mass increases, the planet envelope contracts due to both the stronger gravity and cooling through radiation.
Eventually, when the core mass
reaches $\sim 10 M_{\oplus}$, 
the planet undergoes runaway accretion and contracts significantly. At this stage,
the envelop shrinks significantly and detaches from the planet's Hill sphere (Papaloizou \& Nelson 2005). 
Material
that resides beyond the Hill sphere can still flow into the Hill sphere, but
 forms a circumplanetary disk (CPD) around the protoplanet 
to conserve angular momentum (Lubow \etal 1999; Ayliffe \& Bate 2009). 
The accretion of the circumplanetary disk (CPD) onto the planet allows the continuous growth of the planet even after
 the planet's runaway growth stage. 

{Circumplanetary disks (CPDs) may provide observational signatures of young forming planets in disks.}
Giant planets can be too faint to be detected, but accreting CPDs can be bright and
detectable. For example,
to form a giant planet of 1$-$10 M$_{J}$  mass within the circumstellar disk's life time ($\sim$ a 
few million years, Hernandez \etal 2007), the CPD needs to accrete at a rate of 
$\dot{M}\gtrsim 10^{-9}-10^{-8} \msunyr$. Such an accretion disk will have a 
luminosity (e.g. Owen 2014, Zhu 2015)  of 
\begin{equation}
L_{disk}=\frac{G{ M}_{p}\dot{M}}{2 {\rm R}_{J}}=1.5\times10^{-3}{\rm L}_{\odot} \frac{{M}_{p}}{1 {\rm M}_{J}}\frac{\dot{M}}{10^{-8}\msunyr}\,,
\end{equation}
which is as bright as a late M-type/early L-type brown dwarf and can be detected by current
direct imaging techniques (Zhu 2015). Unlike a planet or brown dwarf which has an almost constant surface
temperature, an accreting CPD has a lower temperature at
larger disk radii, and those outer disk regions will emit significant infrared flux. 
Thus, the emission from an accreting CPD is redder than the emission from a planet or a brown dwarf.  
Direct imaging observations have found several red sources within circumstellar disks (Kraus 
\& Ireland 2012, Quanz \etal 2013, Biller \etal 2014, Reggiani \etal 2014, Currie \etal 2015) and their photometry at 
near-IR bands are consistent with accreting CPDs (Zhu \etal 2015). 
{  $H_{\alpha}$ emission lines, which are another  observational signature of accretion disks, are also found in
some low-mass substellar objects (Zhou \etal 2014, Bowler et al. 2015). }
The most direct evidence for accreting CPDs is LkCa 15b, where two accretion tracers, H$_{\alpha}$ line and near-IR thermal
emission, have both been detected (Sallum \etal 2015). 

CPDs are also essential for satellite formation.
In our solar system, most satellites around giant planets are in prograde, nearly circular and coplanar orbits, implying that they formed in a shared CPD orbiting within the planet's equatorial plane. The  ratio between the total mass of four major Galilean satellites and the Jupiter's mass 
is $\sim 2\times10^{-4}$.
The Saturnian satellite system also has a similar mass fraction with respect to the Saturn's mass (Canup \& Ward 2006). Assuming the gas-to-dust mass ratio is 100, a minimum gas mass
of $\sim 0.02$ planet mass is required to produce these satellites (Canup \& Ward 2002, 2006). There are two main scenarios for Galilean satellite formations. 
One assumes that the satellites form in-situ in a CPD containing this amount of material (e.g. Lunine \& Stevenson 1982; 
Mosqueira \& Estrada 2003). This disk is referred to as the ``minimum 
mass sub-nebula'', which is the analog of ``minimum mass nebula'' for the circumstellar disk. Spreading 0.02 Jupiter mass
within 30 Jupiter radii (where the four major satellites reside), the minimum mass sub-nebula 
has a very high gas surface density $\sim 10^{5} {\rm g}\,{\rm cm}^{-2}$. In the other scenario, the satellites form in a CPD with 
a much lower surface density, but the disk is dynamically evolving and being supplied
by the circumstellar disk  (Canup \& Ward 2002, 2006). 
This ``gas-starved'' scenario only requires that the total supplied mass during the whole satellite formation timescale 
is  larger than $0.02$ planet mass.

The key question for understanding the structure of CPDs is how CPDs accrete. They may accrete in similar ways
as  circumstellar disks accrete.
Turner \etal (2014) suggest that the surface of CPDs can be ionized by X-rays so that the magnetorotational instability (MRI)
can operate at the disk surface leading to accretion, a process similar to the layered accretion proposed 
in protoplanetary disks (Gammie 1996). However, Fujii \etal 
(2011, 2014) find that the active layer is so thin ($\Sigma\sim10^{-3}-10^{-2}{\rm g} \,{\rm cm^{-2}}$) in CPDs that the accretion through 
MRI is negligible. On the other hand, other non-ideal MHD effects (e.g. Hall effects, Kunz \& Lesur 2013, Lesur \etal 2014, Bai 2014) which are important in
protoplanetary disks  can also be
important in CPDs, and CPDs may accrete through magnetic breaking (Keith \& Wardle 2014).   Magnetocentrifugal disk wind can also be launched  in CPDs,  carrying away angular momentum and leading to disk accretion (Quillen \& Trilling 1998, Gressel \etal 2013).
Almost all these proposed CPD accretion mechanisms  (e.g. layered accretion, non-ideal MHD effects, disk wind) can find their roots in circumstellar disk models. Thus,
they are facing the same uncertainties as the accretion mechanisms in circumstellar disks: they sensitively depend on  the net magnetic fields assumed and the detailed microphysics in the disk (e.g. dust
size distribution).

On the other hand, CPDs are different from circumstellar disks in that
they are subject to the tidal torque from the central star, and truncated within the Hill sphere
of the planet (Martin \& Lubow 2011a). In addition,  circumstellar disk material flows
through the Hill sphere, continuously replenishing CPDs. These properties make CPDs similar to disks with inflows from companion stars
in close binary systems or Cataclysmic Variables (CV).
In these binary systems, the tidal torque from the companion star will excite spiral density waves in 
disks, and when these waves shock in disks, they
can transport angular momentum to the disk  leading to disk accretion. 
Previous inviscid isothermal simulations by Rivier \etal (2012) have suggested that accretion due to spiral shocks in CPDs is inefficient  with Jupiter's mass doubling time $\sim$ 5 Myrs.  Szul{\'a}gyi \etal (2014) have measured a much higher accretion rate in their 3-D isothermal inviscid simulations (10$^{-4}$ M$_{J}$/yr), but this measured accretion rate is not numerically converged with their higher resolution simulations. Instead, by  measuring the torque exerted by the star in the simulations, they estimate that the real accretion rate is $\sim 2.5\times 10^{-6}$M$_{J}$/yr.

Recent studies on spiral shocks (Ju \etal 2016)  have shown that the accretion due to spiral shock dissipation  is sensitive to the disk thermodynamics assumed. When the disk is hot and the Mach number, defined as the ratio between the Keplerian speed and the sound speed, is small ($<$10), the equivalent $\alpha$ 
of shock-driven accretion can reach $\sim$0.01-0.02. The Mach number 
of CPDs depends on the disk accretion rate and is sensitive to 
the equation of state applied in simulations (D'Angelo \etal 2003, Ayliffe \& Bate 2009, Machida 2009,  Szul{\'a}gyi \etal 2016). 

Considering the potential importance of spiral shocks driving accretion in CPDs, in this paper
 we construct two-dimensional inviscid hydrodynamical simulations  
to study CPDs. 
These inviscid simulations differ from most previous works that
use artificial viscosity (e.g. $\alpha$ viscosity) to sustain the disk accretion. 
Furthermore, to reduce the numerical viscosity in the CPD region,
we adopt a grid structure centered on the planet. 
Since the thermodynamics which controls the disk Mach number is crucial for the shock-driven accretion, 
a simple radiative cooling scheme has been included in the simulations, which differs this work from inviscid
simulations by Rivier \etal (2012) and Szul{\'a}gyi \etal (2014). We have also measured
the mass accretion rate directly from simulations, unlike these two works where they use the torque exerted by the star onto the CPD to estimate
the disk accretion rate. As Ju \etal (2016)
has pointed out, shock-driven accretion is  determined
by the shock dissipation (the difference between the tidal torque exerted to the disk and the angular momentum flux
carried away by the wave), instead of the total torque alone. We will show that, by allowing the disk being heated up by the shock, the inviscid simulation is numerically converged
and the disk can reach a steady state, transferring inflow material from the Hill radius all the way to the central planet quasi-steadily. 
This shock driven accretion is very efficient with $\alpha\sim 0.001-0.01$.

One caveat in our simulations is that our simulations are limited to 2-D and 
numerous previous 3-D simulations have shown that  the infall from circumstellar to circumplanetary
disks  occurs at high altitudes (Bate \etal 2003, Machida \etal 2008, Tanigawa \etal 2012, 
Morbidelli \etal 2014, Szul{\'a}gyi \etal 2014, 2016).
Furthermore, since the CPD
is significantly heated and puffed up, our simple cooling treatment based
on the thin disk approximation is inaccurate. 3-D simulations with realistic radiative transfer (Szul{\'a}gyi \etal 2016),
 thermodynamics, planet evolution (Ward \& Canup 2010) and even planetesimal accretion (D'Angelo \& Podolak 2015) are 
 needed in future to confirm if spiral shocks can lead to
efficient accretion in CPDs. 

In \S 2,  our numerical method is introduced. The results are presented in \S 3, including both
the disk structure and the accretion rate. 
After a short discussion in \S 4, the paper  will be concluded in \S 5. The detailed energy budget of the
accretion process is given in the appendix.

\section{Method}
We solve the Euler equations to study dynamics in CPDs using Athena++ (Stone \etal 2016, in preparation).
Athena++ is a newly developed grid based code using a higher-order Godunov scheme for MHD and 
the constrained transport (CT) to conserve the divergence-free property for magnetic fields. 
Compared with its predecessor Athena 
(Gardiner \& Stone 2005, 2008; Stone \etal 2008),  Athena++ is highly optimized for speed and uses flexible grid structures, allowing global
numerical simulations spanning a large radial range. Furthermore, the geometric source terms in curvilinear coordinates 
(e.g. in cylindrical and spherical-polar coordinates) are carefully implemented so that angular momentum
is conserved exactly (to machine precision), which is crucial for angular momentum budget analysis in \S 3.2.

In this paper, magnetic fields are ignored. We solve hydrodynamical equations  in 
the cylindrical coordinate system  ($R$ and $\phi$) that is centered
on the planet. This grid choice is different from most previous works which 
have adopted cylindrical or spherical coordinate systems
that are centered on the star. Having the cylindrical coordinate system centered on the planet
allows the material in CPDs flow along the azimuthal grid direction, thus significantly reduces numerical truncation errors. 
Adopting the cylindrical coordinate system with uniform grids in ln($R$) also allows each grid to have the same
length along the azimuthal and radial direction throughout the whole simulation domain .

Furthermore, we adopt the rotating frame so that the star is stationary at $R$=1, $\phi$=0. 
The equations being solved are
\begin{align}
\frac{\partial \Sigma}{\partial t}+\nabla\cdot(\Sigma \mathbf{v})&=0,\\
\frac{\partial \Sigma\mathbf{v}}{\partial t}+\nabla\cdot(\Sigma \mathbf{v}\mathbf{v}+P \mathbf{I})&=-\Sigma\nabla\Phi_{p}+\mathbf{F},\label{eq:vphi}\\
\frac{\partial E}{\partial t}+\nabla\cdot[(E+P)\mathbf{v}]&=-\Sigma \nabla\Phi_{p}\cdot\mathbf{v}+\mathbf{F}\cdot\mathbf{v}-Q_{c}
\end{align}
where $\Sigma$ is the disk surface density, $\Sigma \mathbf{v}$ is the linear momentum, $\Phi_{p}$=-GM$_{p}$/R 
is the potential of the planet having the mass of M$_{p}$, and
 $E=\epsilon+\frac{1}{2}\Sigma v^{2}$ is the total energy density per unit area with $\epsilon$ being the internal energy.
 Note that since the planet is at the coordinate center, coordinate $R$ and $\phi$ are the position  
 relative to the planet instead of
 the star. $\mathbf{F}$ is the total external force excluding
 the gravitational force from the planet, and thus  $\mathbf{F}$ includes the
centrifugal force and Coriolis force due to the adoption of the rotating frame, the direct gravitational force from the star, and 
the indirect force (since the grid is centered at the planet instead of the center of mass). The detailed forms of these terms
are given in Ju \etal (2016). Solving these equations in cylindrical
coordinates also introduces geometrical source terms, which are written in forms of conserving the angular momentum with machine error precision (as in Ju \etal 2016). As will be shown in \S 3.2, 
maintaining angular momentum conservation is crucial for the  angular momentum diagnostics.  

No physical viscosity has been applied in our simulations. Thus, angular momentum and energy transport can only come from
shock dissipation and numerical viscosity. Since simulations having twice resolution produce almost identical results, which will be shown below, 
we conclude that numerical viscosity is negligible and angular momentum and energy transport is dominated by the shock dissipation. 

The equation of state for an ideal gas has been assumed, so that $P=\epsilon(\gamma-1)$. We further assume  $\gamma$=7/5. The
  density-weighted  disk temperature is  $T=\mu P/(\Re \Sigma)$, where $\Re$ is the gas constant and $\mu=2.4$.
  Since most disk mass is concentrated at the disk midplane, we use $T$ to approximate  the disk midplane temperature $T_{c}$. 

The cooling rate per unit area, $Q_{c}$, is approximated by
\begin{equation}
Q_{c}=\frac{16}{3}\sigma(T_{c}^{4}-T_{\rm ext}^{4})\frac{\tau}{1+\tau^{2}}\,,
\label{eq:cooling}
\end{equation}
where $\tau=(\Sigma/2)\kappa_{R}(\rho_{c},T_{c})$ is the optical depth to the 
disk midplane at radius $R$ (Hubeny 1990). The midplane density $\rho_{c}$ is $\Sigma/(\sqrt{2\pi} H)$ where $H$ is the disk scale height $c_{s}(T_{c})/\Omega$.  The Rosseland mean opacity $\kappa_{R}(\rho_{c},T_{c})$ is assumed to be
10 ${\rm cm^{2} g^{-1}}$, which is the typical value (D'Alessio \etal 2001) at  temperatures below the dust sublimation temperature of $\sim$1500 K \footnote{In some of the runs,
the temperature of the very inner disk goes above the dust sublimation temperature and the opacity should
decrease dramatically. However, we still use a constant opacity to avoid disk instability associated with the opacity change (Zhu \etal 2009).  }. The
particular form $\tau/(1+\tau^{2})$ is chosen so that the cooling
term has the correct form in both optically thick and thin limits \footnote{ 
In the optically thin limit, we  should use the Planck mean opacity.  
Because the dominant opacity is from dust, which has a relatively slow variation with 
wavelength, we use the Rosseland mean opacity for simplicity.}.
The term $\sigma T_{\rm ext}^{4}$ represents the energy flux into the disk from the stellar irradiation,  
leading to a background disk temperature of
 $T_{\rm ext}$.
 
Due to the inclusion of the thermodynamics in the problem, we have to specify physical
units in the simulations. We assume that the distance between the star and the planet is
5 AU,  the star is a solar mass star, and the planet's mass is 0.001 $M_{\odot}$ ($M_{J}$). 
$T_{\rm ext}$ is 128 K so that the aspect ratio of the circumstellar disk at the planet position  ($c_{s}/v_{\phi}$ where $v_{\phi}$ is Jupiter's orbital velocity around the Sun) is 
0.05. Besides the simulations with radiative cooling, we have also carried out simulations with isothermal
and adiabatic equations of state for comparison.

\begin{figure*}[ht!]
\centering
\includegraphics[trim=0cm 0cm 0cm 0cm, width=0.9\textwidth]{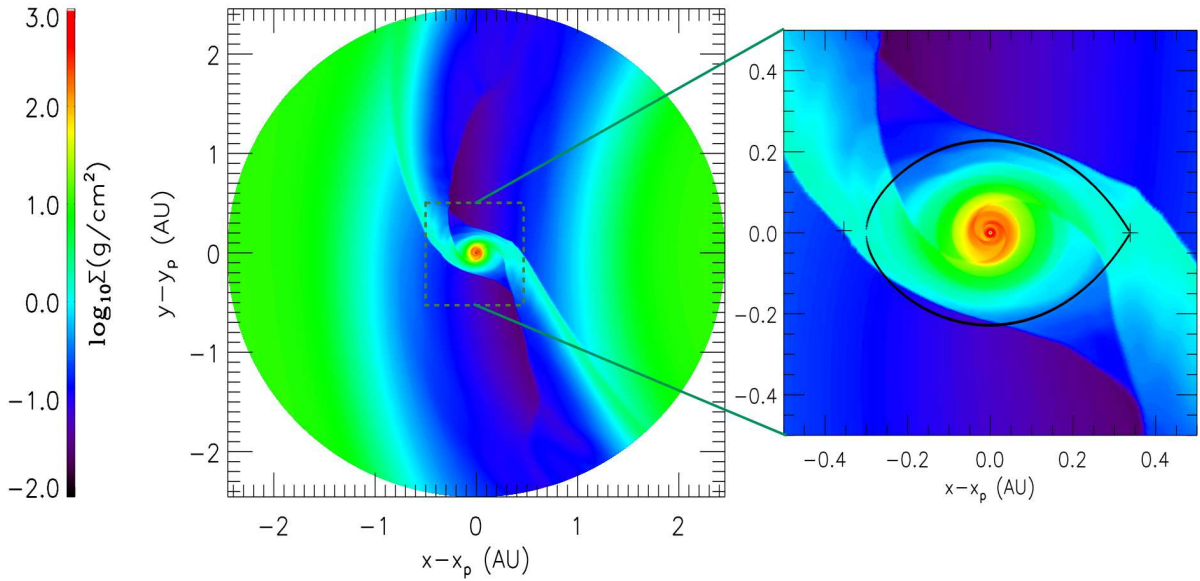} 
\caption{ The disk surface density  at t=100 $\Omega_{5 AU}^{-1}$ for the case Gp3. The right panel zooms in the CPD region. The two crosses in the right panel are L1 and L2 points of the planet while the black curve represents the Roche lobe of the planet. }
\vspace{-0.1 cm} \label{fig:Surf}
\end{figure*}

Our cylindrical grids are  uniformly 
spaced from $0$ to $2\pi$ in the  $\phi$ direction, and uniformly
spaced in ln($R$) from $R=$2 or 4 Jupiter radii to half the distance between the planet and the star (which is 2.5 AU). Given an adopted resolution in the radial direction,  we adjust the resolution in the azimuthal direction 
to make sure every grid cell is square in physical size. Since a Jupiter mass planet should induce a gaseous gap
in the circumstellar disk (Lin \& Papaloizou 1986, Kley \& Nelson 2012), we have manually carved out a gap in the simulations as shown
in Figure \ref{fig:Surf}. Motivated by detailed gap opening studies (e.g. de Val-Borro \etal 2006, Fung \etal 2014), we parameterize
the gap density profile as
\begin{equation}
\Sigma(\mathbf{r}) =
  \begin{cases}
    \Sigma_{l}+\frac{\Sigma_{h}-\Sigma_{l}}{2}\times\left(2-{\rm exp}\left\{\frac{-\Delta R_{dp}+w_{g}}{w_{t}}\right\}\right)      \\
    \quad \text{if } \Delta R_{dp}>w_{g}\\
    \\
    \Sigma_{l}+\frac{\Sigma_{h}-\Sigma_{l}}{2}\times\left({\rm exp}\left\{\frac{\Delta R_{dp}-w_{g}}{w_{t}}\right\}\right)      \\
    \quad \text{if } \Delta R_{dp}<w_{g}\\
  \end{cases}
\end{equation}
where $\Sigma_{l}$ and $\Sigma_{h}$ are the disk surface density within and outside the gap, 
$w_{g}$ is the gap half width, $w_{t}$ represents the sharpness of the gap, and
 $\Delta R_{dp}\equiv |R_{*}-|\mathbf{R}-\mathbf{R_{*}}||$ represents how far away each point is from
 the gap center, and $\mathbf{R_{*}}$
is the star's position in the planet centered frame. In all our simulations, we choose $w_{t}=0.25$ AU, which is roughly
1/6 to 1/4 of the gap half width. 
The velocity field is initialized by maintaining hydrostatic equilibrium in the radial direction by balancing
 the pressure gradient, the central stellar gravity, centrifugal and Coriolis forces.  We have tested that
  the gap shape is nicely maintained in the simulation where there is no planet at the center.
 We also initialize a Keplerian
 CPD with a very low surface density ($\sim 10^{-4}$ g cm$^{-2}$) within 0.4 Jupiter's Hill radii.

To maintain the gap structure during the simulation, we fix physical quantities at the outer boundary to be the initial values. 
At the inner boundary, we adopt the ``outflow, no inflow'' open boundary condition. The density and radial velocity are copied
from the last active zones to the ghost zones. When the radial velocity is towards the computational domain, it is set to be zero.
The azimuthal velocity in the ghost zones are set to their local Keplerian values. Internal energy in the ghost zones is the same
as that in the last active zones. 

Our initial and boundary condition are quite different from those used in previous circumplanetary disk simulations. 
Instead of capturing the planet gap opening process in the simulation, 
we prescribe the gap density profile based on detailed gap opening studies (e.g. de Val-Borro \etal 2006, Fung \etal 2014).
By this approach, the gap shape is fixed and the inflow from the circumstellar disk to the circumplanetary disk can reach
a steady state. In reality, the detailed gap width and depth depend on the disk turbulent level, the disk thermal structure
and even the magnetic field geometry (Zhu \etal 2013). In inviscid disks, the gap will even become deeper with time (Zhu \etal 2013).
To simplify the gap opening process,
we use different gap widths and depths in different simulations,  to represent the gaps in disks with different properties 
or during different stages of the gap opening process. 
We also use these different gap widths and depths to control the inflow rate to the CPD. In our simulations, 
we allow the circumplanetary region to settle down by itself and have no direct control of the inflow rate to the CPD.
Thus, we take advantage of the fact that the simulation with a wider and deeper gap has a
lower inflow rate to the CPD, and, through varying gap shapes, we indirectly control the inflow rate to cover 4 orders of magnitude.

\begin{table}[ht]
\begin{center}
\caption{Models with Prescribed Gap Structure \label{tab1}}
\begin{tabular}{cccccccc}
\tableline
\tableline
\tableline
Run    &  EoS & Resolution & $\Sigma_{h}^{\,\,\,\,\, b}$  & $\Sigma_{l}$ & $w_{g}$ & Time   \\
           &   & $R\times\phi$ & g cm$^{-2}$  & g cm$^{-2}$  & $AU$ & yr    \\
\tableline
Gp2  & Cool &432$\times$352   & 10  & 0.1 &  1 & 210 \\
Gp2H  & Cool &864$\times$704   & 10  & 0.1 &  1 & 85 \\
Gp2adi  & Adi. &432$\times$352   & 10  & 0.1 &  1 & 250 \\
Gp3 & Cool &432$\times$352   & 10  & 0.1 &  1.5 & 250 \\
Gp3iso & Iso. &432$\times$352   & 10  & 0.1 &  1.5 & 297 \\
Gp3Sp1 & Cool &432$\times$352   & 1  & 0.01 &  1.5 & 515 \\
Gp3Sp01$^{a}$ & Cool &400$\times$352   & 0.1  & 10$^{-3}$ &  1.5 & 1216 \\
Gp3Sp001$^{a}$ & Cool &400$\times$352   & 0.01  & 10$^{-4}$ &  1.5 & 2485 \\
\tableline
\end{tabular}
\tablenotetext{1}{ The inner boundary of the simulation domain is  at 4 Jupiter radii instead of 2 Jupiter radii.
The initial surface density of the CPD is set to be 10 g cm$^{-2}$, and the mass quickly drains to the planet at the early stages.}
\tablenotetext{2}{ $\Sigma_{h}$ and $\Sigma_{l}$ are the disk surface density  outside and inside the gap, and
$w_{g}$ is the gap half width.}
\end{center}
\end{table}

All simulations are summarized in Table 1. The number after the capital letter ``G'' (where p2 means 0.2) represents
the gap half width ($w_{g}$) in the code unit (1 is 5 AU). The number after  the capital letter S represents the circumstellar
disk surface density ($\Sigma_{h}$) relative to the surface density in the fiducial case (Gp2). 
Our fiducial case (Gp2) has a circumstellar disk with
10 g cm$^{-2}$ surface density, which is one order of magnitude smaller than 
the surface density at 5 AU based on the minimum
mass solar nebula model (Hayashi \etal 1981). The simulations having ``iso'' and ``adi'' in their names are isothermal and adiabatic simulations.
All other simulations have considered simple radiative cooling (Equation \ref{eq:cooling}). 
For simulations with low circumstellar surface density and thus 
low inflow rates (Gp3Sp01 and Gp3Sp001), the CPDs take much longer to a reach steady state. Thus,
we have to adopt a bigger inner boundary (4 Jupiter radii) to speed up the calculations
and a higher initial CPD surface density to reduce the time to reach a steady state.

\section{Results}
The typical disk surface density during the simulation is shown in Figure \ref{fig:Surf}. The morphology of CPD is very similar to
previous studies (e.g. Lubow \etal 1999). 
Gas material in the circumstellar disk enters the Roche sphere of the planet when the gas   
 undergoes horseshoe orbits around the planet.
The planet excites spiral arms in the circumstellar disk, while the star excites spiral arms in the CPD. These
spiral arms in CPDs are crucial for angular momentum transport in CPDs as shown in \S 3.2.

\subsection{Disk Structure}
\begin{figure*}[ht!]
\centering
\includegraphics[trim=0cm 0cm 0cm 0cm, width=0.48\textwidth]{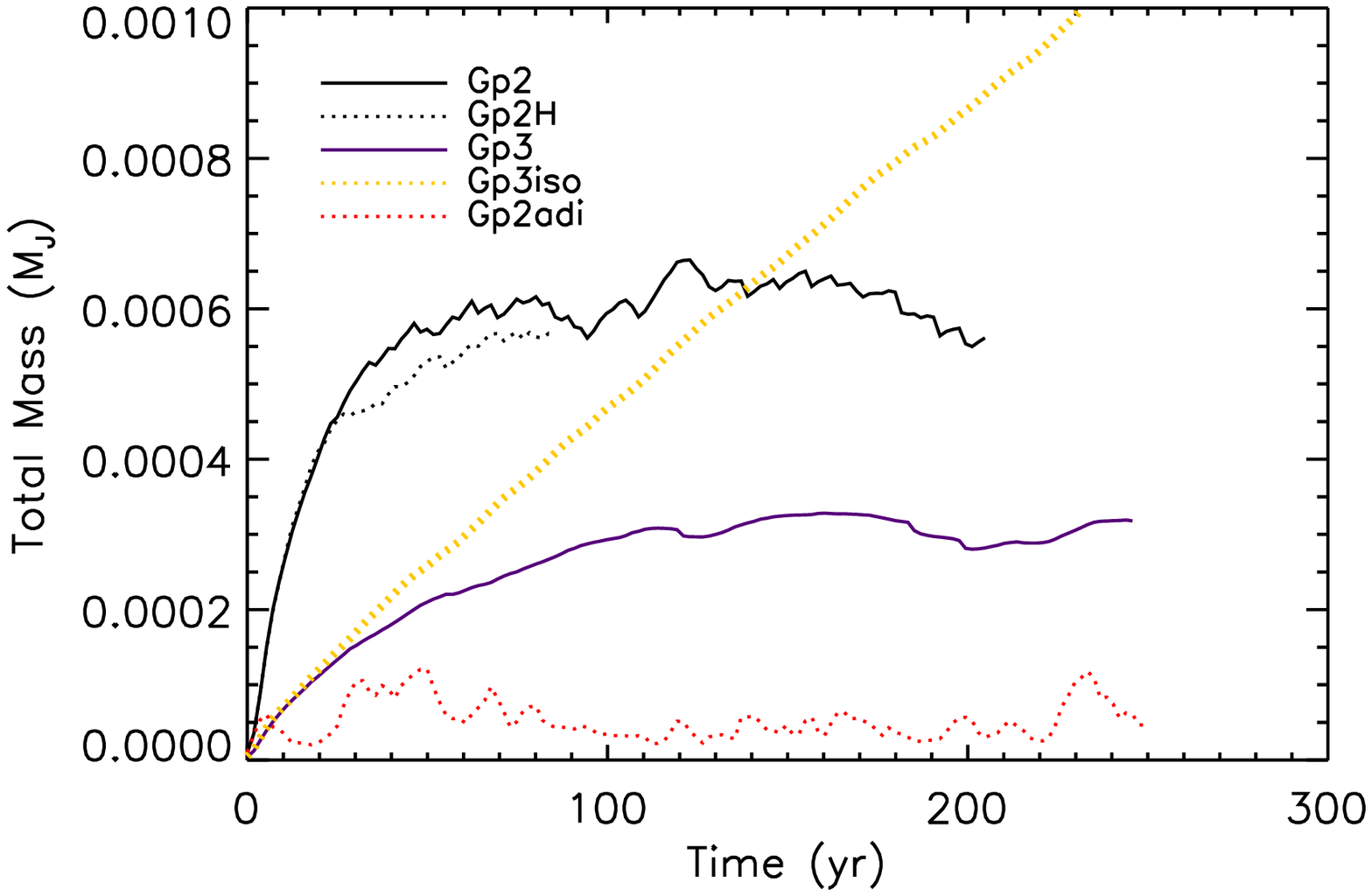} 
\includegraphics[trim=0cm 0cm 0cm 0cm, width=0.48\textwidth]{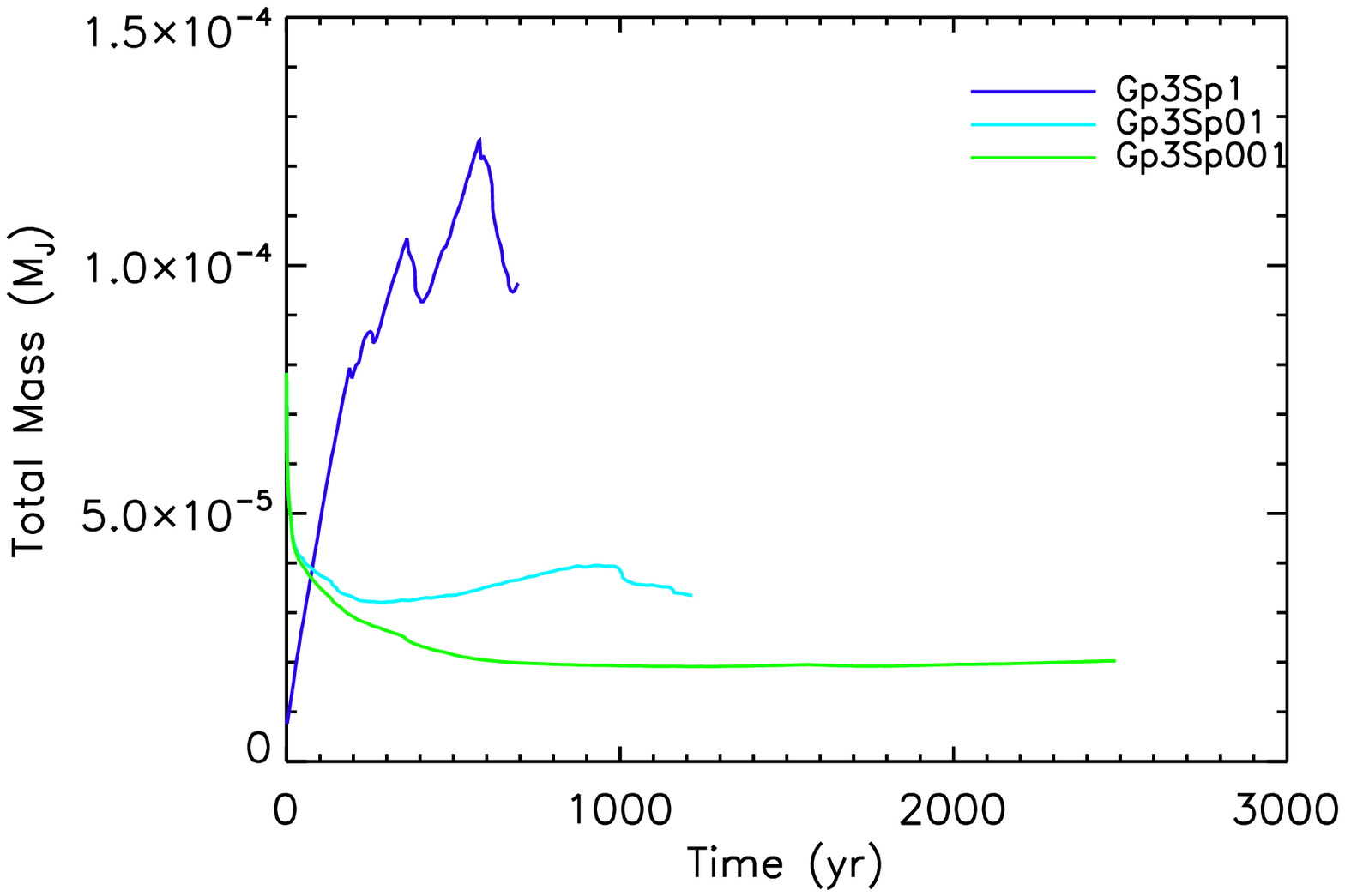} 
\caption{ The mass of CPDs with time. All cases reach quasi-steady state except for the isothermal case (the yellow dotted curve).
}
\vspace{-0.1 cm} \label{fig:masstime}
\end{figure*}

We have run these simulations until they reach quasi-steady state, except for the isothermal case which shows
no sign of reaching  a steady state during the whole run \footnote{After a long simulation time, the isothermal case may eventually reach a steady state when enough material is piled up in the CPD. }. The total mass of the CPD within the Hill sphere ($R<$0.347 AU)
 is shown in Figure \ref{fig:masstime}.  When a quasi-steady state has been reached, the accretion rate onto the planet 
 equals the inflow rate, and the disk mass remains a constant. Disks with narrower 
 (smaller $w_{g}$) gaps and higher surface density 
 (higher $\Sigma_{h}$ and $\Sigma_{l}$) have higher inflow rate from the circumstellar disk to the CPD.
 For case Gp2 which has the highest accretion rate, the CPD reaches a steady state
 within 50 years. 
 For cases with low accretion rates (e.g. Gp3Sp001), the disk reaches  a quasi-steady state on the timescale of 
1000 years. For a comparison with the dynamical timescale, the orbital time at our inner boundary (2 R$_{J}$) 
and half the Hill radii is only 8 hours and 2.3 years respectively. Thus, 100 years which is the typical duration of the simulation  is 1.1$\times 10^{5}$ innermost orbits. 

\begin{figure*}[ht!]
\centering
\includegraphics[trim=0cm 0cm 0cm 0cm, width=0.9\textwidth]{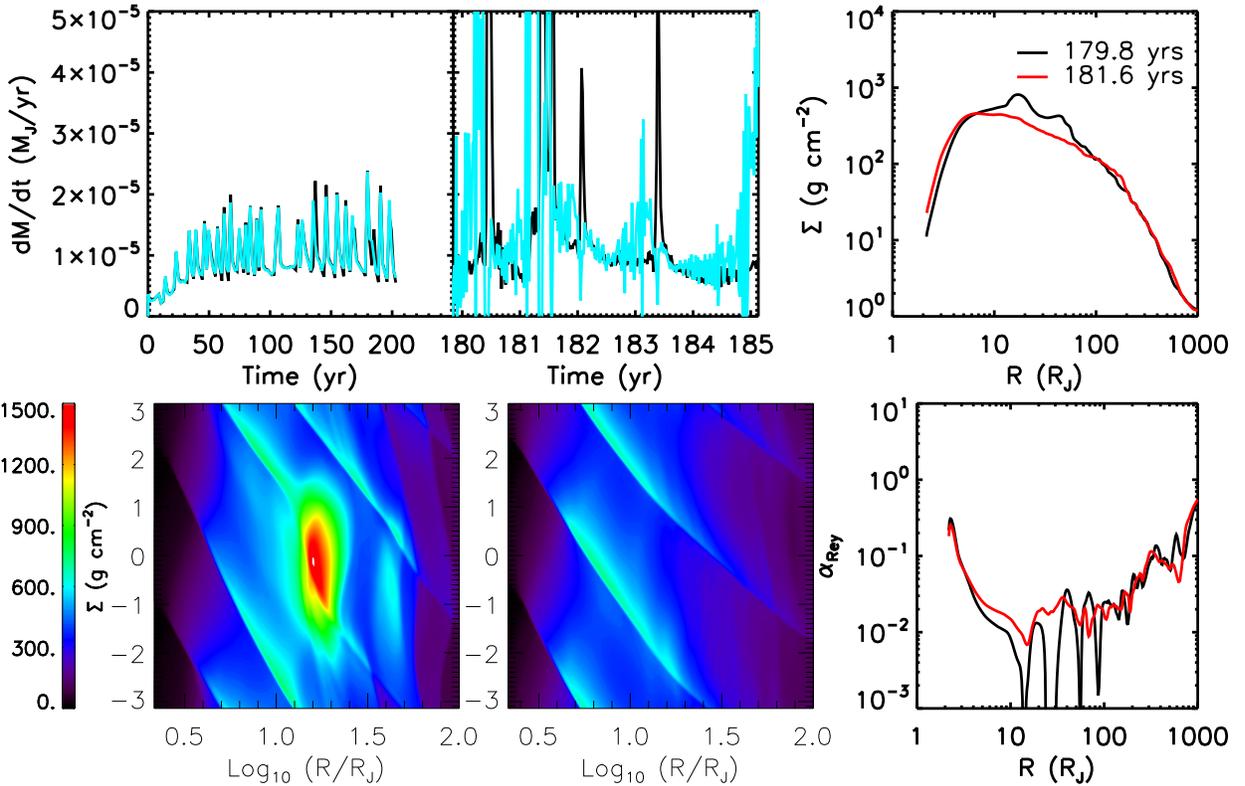} 
\caption{ Upper left panel: the disk accretion rates that are averaged over every $\Omega_{5 AU}^{-1}$ at the inner boundary (black curves) and at $R=10 R_{J}$ (blue curves) for Gp2. Upper middle panel: similar to the left panel but the accretion rates are averaged over every $0.01\times \Omega_{5 AU}^{-1}$.
Upper right panel: the disk surface density profile at two different times during one major accretion peak. 
Lower left two panels: the disk surface density contour at these two times (the left is before and the middle is during
the accretion peak.). Lower right panel: the profile of the normalized Reynolds Stress at these two times. }
\vspace{-0.1 cm} \label{fig:vary1}
\end{figure*}

\begin{figure*}[ht!]
\centering
\includegraphics[trim=0cm 0cm 0cm 0cm, width=0.9\textwidth]{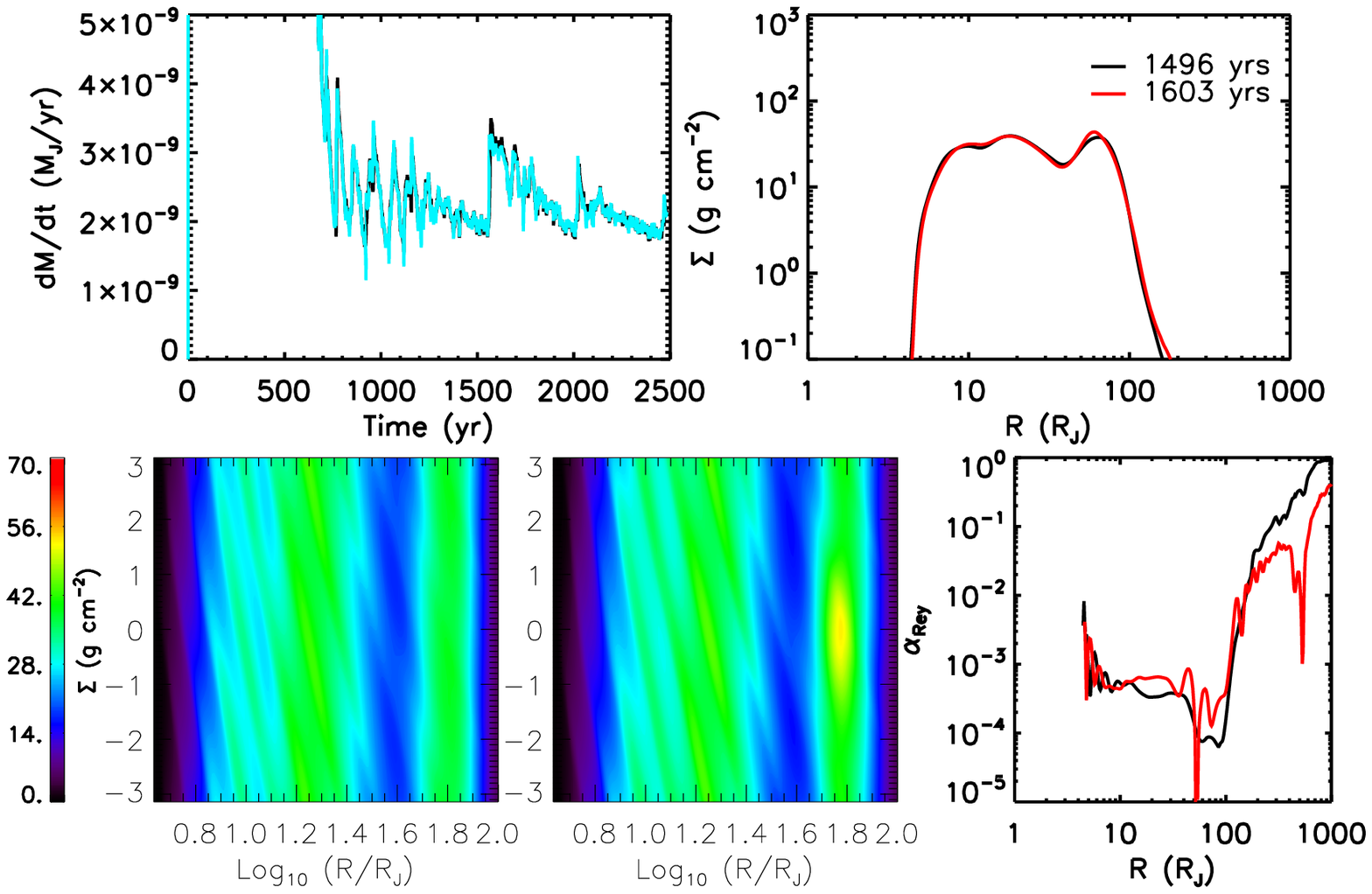} 
\caption{ Similar to Figure \ref{fig:vary1} but for Gp3Sp001. }
\vspace{-0.1 cm} \label{fig:vary2}
\end{figure*}

The accretion onto the planet is not strictly steady with time. As shown in Figures \ref{fig:masstime}-\ref{fig:vary2} and
also in Ju \etal (2016), there 
are short timescale variabilities. The disk accretion rates at the inner boundary and at $R$=10 $R_{J}$ are
shown in the upper left panels of Figure \ref{fig:vary1} and \ref{fig:vary2}  for both Gp2 and Gp3Sp001 respectively. The accretion rates
in these panels have been averaged over 1 $\Omega_{5AU}^{-1}$. For case Gp2, there is significant variability even within the timescale
of 1 $\Omega_{5AU}^{-1}$. Thus, we plot the accretion rates at a much higher cadence (averaged over 0.01 $\Omega_{5AU}^{-1}$) within a short period of time.
To understand the short-timescale variability, we plot
the disk surface density profile at different times during the outburst, as shown in the upper right panels of Figure \ref{fig:vary1} and \ref{fig:vary2}. {  The 2-D surface density contours at these two times are shown in the lower left two panels of these figures (the lower left panel corresponds to the black curve in the upper right panel, while the lower middle panel corresponds to the red curve.). 
We can see that  the disk develops significant asymmetric 
structure during these accretion peaks.} Vortices are generated in these disks first. As clearly shown
in Figure \ref{fig:vary2}, banded
structures are first induced by spiral shocks, and these structures are subject to Rossby Wave Instability (Lovelace \etal 1999) and become vortices. {  In the lower left panel of Figure \ref{fig:vary1}, a vortex forms at the outer disk and migrates inwards. Eventually it crosses the inner boundary (lower middle panel), leading to an outburst. In Figure \ref{fig:vary2}, the vortex does not migrate significantly in the disk, but it excites spiral arms, leading to additional angular momentum transport.} 

{  Thus, vortices provide some other ways to transport angular momentum in the disk, which are different from the accretion process 
due to star-induced spiral shocks mentioned above. First,
vortices can excite spiral density waves, which can become spiral shocks and transport angular momentum. Secondly, 
vortices themselves
will  migrate in disks  (Paardekooper \etal 2010), and this vortex migration directly leads to mass transport since 
the vortex region has a high disk surface density.
Thus, it is important
to know how much disk accretion is driven by star-induced spiral shocks and how much accretion is due to vortices generated. Unfortunately,
using Reynolds stress alone, as in \S 3.2, we cannot distinguish these accretion mechanisms. On the other hand, 
we can use the change of the Reynolds stress when a vortex appears or disappears in the disk to qualitatively study the role
played by the vortex. As shown in the bottom right panels of Figures \ref{fig:vary1} and \ref{fig:vary2}, the Reynolds stress does not change significantly when there is a vortex in the disk.  This suggests that the spiral shocks play a more important, if not dominate, role
in transporting angular momentum in disks.}

{  On the other hand, vortices regulate the disk accretion rate, causing episodic accretion. 
And, for the disks with higher accretion rates, vortices 
are generated at smaller radii (by comparing Figure  \ref{fig:vary1} and \ref{fig:vary2}), leading to  shorter timescale and more violent outbursts. }

\begin{figure*}[ht!]
\centering
\includegraphics[trim=0cm 0cm 0cm 0cm, width=0.9\textwidth]{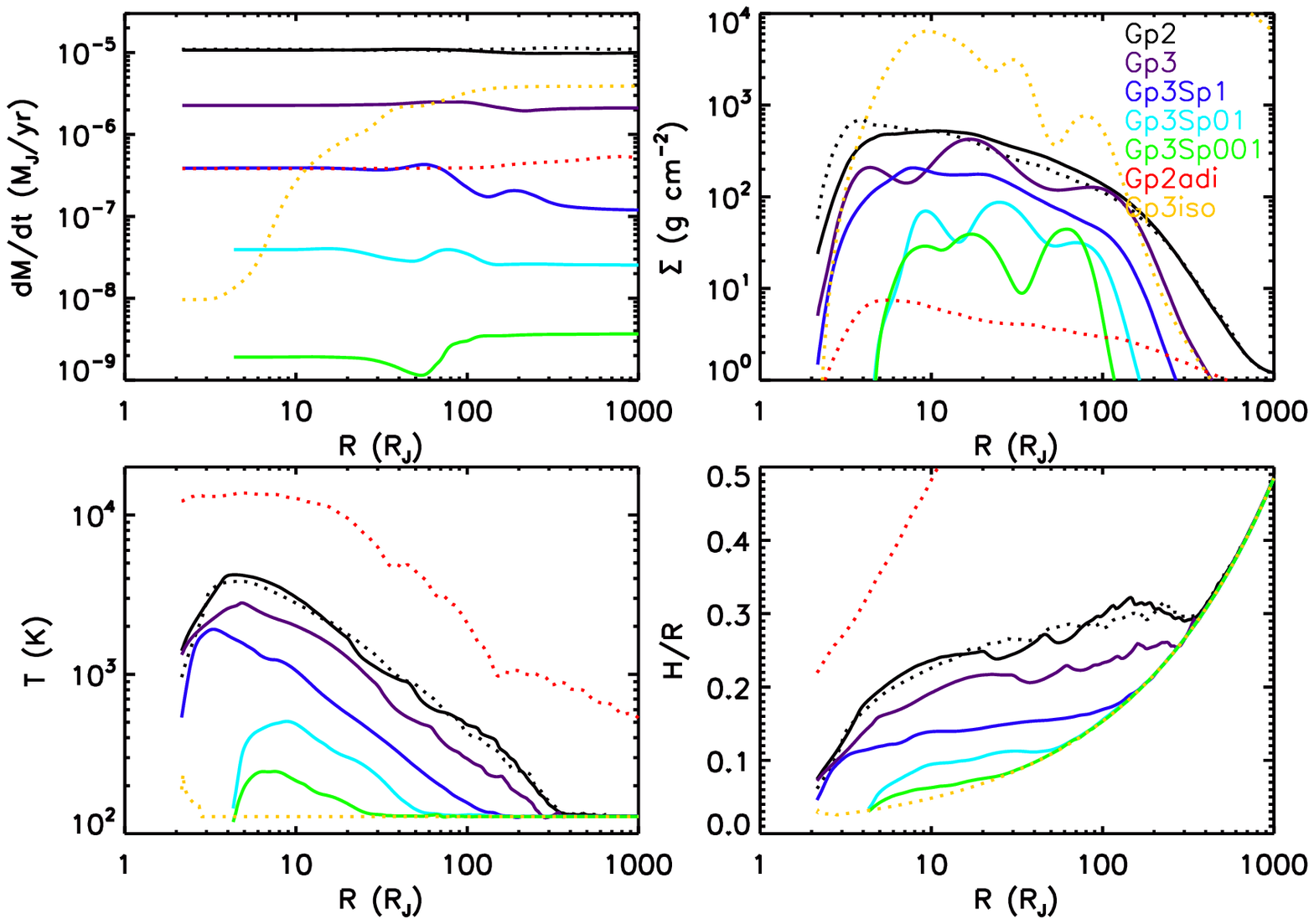} 
\caption{ The CPD mass accretion rate, surface density, midplane temperature and aspect ratio with the radius for all runs. The black dotted
curves are from the high resolution run (Gp2H). The quantities in the upper panels are averaged over each time step for a period of time,
as described in the text. All quantities are azimuthally averaged.}
\vspace{-0.1 cm} \label{fig:radial}
\end{figure*}

To remove the short-timescale variability and study the general properties of disk structure, 
we average the mass accretion rate and surface density at each time step over 
10 $\Omega_{5AU}^{-1}$ (18 years, for the high resolution case, Gp2H), 50 $\Omega_{5AU}^{-1}$ (90 years, for fiducial resolution cases with 2 R$_{J}$ inner boundary), 
or 100 $\Omega_{5AU}^{-1}$ (180 years, for fiducial resolution cases with 4 R$_{J}$ inner boundary) before the end of the simulations, and plot the azimuthally
averaged values in Figure \ref{fig:radial}.  For reference, the Hill radius is $\sim$740 R$_{J}$. 
The time-averaged disk accretion rates in our models span a large range from 10$^{-9}$ M$_{J}$/yr to 10$^{-5}$  M$_{J}$/yr
(the upper left panel). In all the cases except the isothermal case, the accretion rate is almost a constant throughout the whole disk, 
implying that the disk reaches a quasi-steady state. In the isothermal case, the disk has a low temperature and the spiral shocks
are so tightly wound up at the inner disk and spiral shocks are inefficient to carry away angular momentum there (Ju \etal 2016,  also in \S 3.2).  Among all the  models with radiative cooling,
the disk surface density only spans one order of magnitude despite the large range of disk accretion rates.
The maximum surface density is between 50 g cm$^{-2}$ and 500 g cm$^{-2}$ in these models (the upper right panel).  

The disk midplane temperature and aspect ratio at the end of the simulations are shown in the bottom panels of Figure \ref{fig:radial}. 
The aspect ratio of the CPD is defined similarly as  the aspect ratio of the circumstellar disk. It is the ratio between the disk scale height (H) and the radius (R) in the CPD, which is equal to
$c_{s}/v_{\phi}$ with $v_{\phi}$ being the orbital velocity around the planet. 
The disk midplane temperature varies significantly among cases.
In simulations that have realistic cooling, 
the maximum disk temperature ranges from 200 K to 5000 K. The disk  with a higher accretion rate has a higher temperature. 
Due to the large temperature range in different cases,
the disk aspect ratio also varies significantly, from 0.05 in the cases with low accretion rates to 0.2 in the cases with high accretion rates.

Both the isothermal and adiabatic cases are significantly different from other cases with realistic radiative cooling. The isothermal case has not reached a steady state. The mass is continuously being piled up in the disk. The adiabatic case has a too high temperature with $H/R$ larger than 0.5 at $R>10 R_{J}$ so that the disk cooling treatment is inaccurate (Equation \ref{eq:cooling}). The disk is so hot that it also limits the inflow rate to the CPD. 
The infall rate in the adiabatic case is more than one order of magnitude smaller than the infall rate in the isothermal case.  After normalizing the accretion rate with the disk surface density,  $\alpha$ is, however, larger in the adiabatic case than the isothermal case since shocks can propagate to the inner disk more easily.

\subsection{Accretion Mechanism}
\begin{figure*}[ht!]
\centering
\includegraphics[trim=0cm 0cm 0cm 0cm, width=0.9\textwidth]{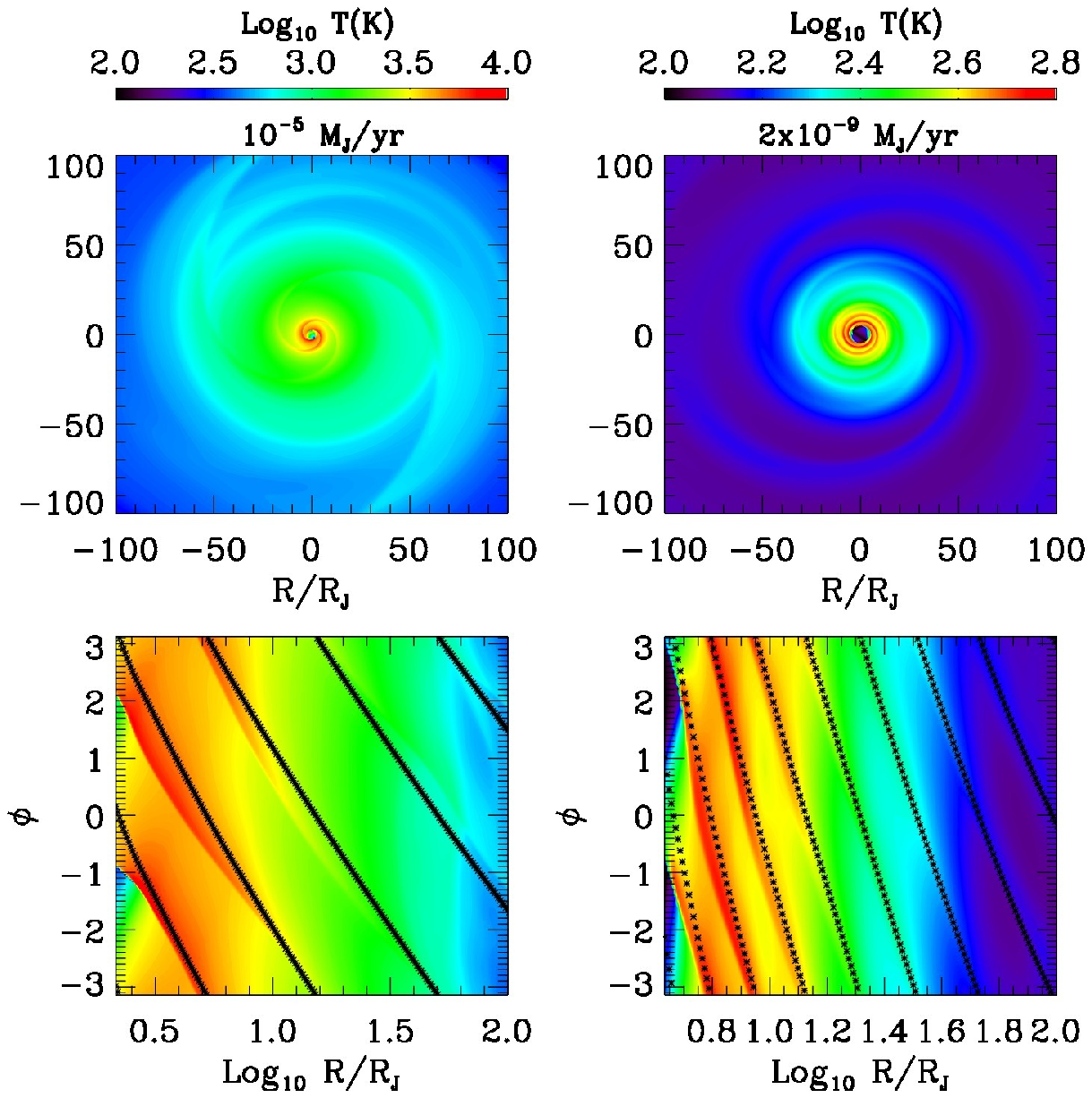} 
\caption{ The temperature color images in the x-y plane (upper panels) and $R-\phi$ plane (bottom panels) for Gp2 (left panels)
and Gp3Sp001 (right panels) at 100 $\Omega_{5AU}^{-1}$. The dotted curves in the bottom panels are spiral arms calculated from linear dispersion relation. }
\vspace{-0.1 cm} \label{fig:spiralT}
\end{figure*}

In our inviscid hydrodynamical simulations,  dissipation of spiral shocks is the main angular momentum transport mechanism. To
show how spiral shocks propagate in CPDs, we plot the temperature color contours from two simulations (Gp2 and Gp3Sp001) with dramatically different disk accretion rates in Figure \ref{fig:spiralT}. 
Two spiral arms are apparent in the figure.  Spiral waves are excited by the tidal force at Lindblad resonances, $R_{Lind}=R_{0} (1\pm 1/m)^{2/3}(1+q)^{-1/3}$ where
$R_{0}$ is the separation between the perturber and the central source, $m$ is the order of the resonance, and q is the mass ratio between the perturber and the central source. 
In our case with the star being the perturber, $q$ is 1000 so that only the inner $m=2$ Lindblad resonance at $R_{Lind}= (1/4004)^{1/3}R_{0}$ is within the Hill radius of the 
planet  $(1/3000)^{1/3}R_{0}$, which is why two spiral arms are excited in Figure  \ref{fig:spiralT}. The strength of the excited spiral waves is strongly correlated with
the disk surface density at Lindblad resonances. After spiral waves are excited, they will propagate to the inner disk. 
During the propagation, linear spiral waves can quickly steepen into spiral shocks  (Goodman \& Rafikov 2001) and dissipate. On the other hand,
these spiral shocks are rather weak and the shock front only deviates slightly from the spiral wave front calculated by the linear dispersion relation (Zhu \etal 2015).
As in Ju \etal (2016), we plot the wave front  by integrating the linear dispersion relation
\begin{equation}
\frac{d\phi}{dR}=-\frac{1}{c_{s,a}(R)}\sqrt{(\Omega-\Omega_{p})^2-\hat{\kappa}^2/m^2}\,.\label{eq:dphidR}
\end{equation}
in the bottom panels of Figure \ref{fig:spiralT} as the dotted curves, where $\hat{\kappa}$ is the epicyclic frequency and $\hat{\kappa}=\Omega$ in a Keplerian rotating disk. $m$ is assumed to be 2, and $c_{s,a}(R)$ is the sound speed 
calculated using the adiabatic equation of state and the azimuthally averaged temperature at each $R$ 
\footnote{Since the disk is highly optically thick, the real sound speed is more close 
to the sound speed calculated with the adiabatic equation of state than the isothermal equation of state. }. 
Figure \ref{fig:spiralT} clearly shows that the linear dispersion relation reproduces the spiral shock front very well. 

The openness of the spiral
arm reflects the disk temperature and also relates to the efficiency of angular momentum transport by spiral shocks (Ju \etal 2016).
This is clearly demonstrated in Figure  \ref{fig:spiralT}. 
The slope of the curves in the bottom panels of Figure  \ref{fig:spiralT}
directly relates to the disk pitch angle, 
\begin{equation}
{\rm cot} \theta=\left|\frac{Rd\phi}{dR}\right|=\left|\frac{d\phi}{d{\rm ln}R}\right|\,.\label{eq:cot}
\end{equation}
With Equations \ref{eq:dphidR} and \ref{eq:cot}, we see that tan$\theta\propto c_{s,a}$. Thus, spiral arms in a hot disk (the left panels in Figure  \ref{fig:spiralT}) have
larger pitch angles and are more open than spiral arms in a cold disk (the right panels). When the spiral arm is more open, 
it is easier to propagate to the inner disk and should lead to 
stronger angular momentum transport there (Ju \etal 2016). Thus, we expect a more efficient angular 
momentum transport and higher $\alpha$ value in the left panels than in the right panels.

To understand the accretion efficiency, we follow Ju \etal (2016)
to separate angular momentum budget in the angular momentum equation. 
The derivations below are basically the same as Shakura \& Sunyaev (1973) and Balbus \& Hawley (1998),
but we wrote them in quantities that can be directly calculated from numerical simulations. 
After being averaged over the azimuthal direction, the angular momentum equation is reduced to
\begin{equation}
\partial_{t}\langle \Sigma R v_{\phi}\rangle _{\phi}=-\frac{1}{R}\partial_{R}(R^{2}\langle \Sigma v_{R}v_{\phi}\rangle _{\phi})+\langle \mathbf{R}\times \mathbf{F}\rangle _{\phi}
\end{equation}
where $\langle X\rangle _{\phi}$ represents $\int_{0}^{2\pi} X  d\phi$, and $\mathbf{F}$ is the external force as in Equation \ref{eq:vphi}.
After subtracting the Keplerian motion of the disk with $\delta v_{\phi}=v_{\phi}-v_{K}$, we have
\begin{align}
&\partial_{t}\langle \Sigma\rangle _{\phi}(Rv_{K}) + \partial_{t}\langle \Sigma R \delta v_{\phi}\rangle _{\phi}\nonumber\\
=&-\frac{1}{R}\partial_{R}\langle R\Sigma v_{R}\rangle _{\phi} (Rv_{K}) -\frac{1}{R}\langle \Sigma R v_{R}\rangle _{\phi}\partial_{R}
 (Rv_{K})\nonumber\\
 &-\frac{1}{R}\partial_{R}(R^{2}\langle \Sigma v_{R}\delta v_{\phi}\rangle _{\phi})+\langle \mathbf{R}\times \mathbf{F}\rangle _{\phi}
\end{align}
After canceling out the first term on the left and right side using the mass continuity equation, we have 
\begin{align}
\partial_{t}\langle \Sigma R \delta v_{\phi}\rangle _{\phi}=&-\frac{1}{R}\langle \Sigma R v_{R}\rangle _{\phi}\partial_{R} (Rv_{K})\nonumber\\
 &-\frac{1}{R}\partial_{R}(R^{2}\langle \Sigma v_{R}\delta v_{\phi}\rangle _{\phi})+\langle \mathbf{R}\times \mathbf{F}\rangle _{\phi}\label{eq:ang}
\end{align}
The left hand is the time derivative of the perturbed angular momentum. To be consistent with Ju \etal (2016), we multiply this quantity by $R^{2}$
and define it as
\begin{equation}
AM_{t}(R)=R^{2} \partial_{t}\langle \Sigma R \delta v_{\phi}\rangle _{\phi} \,.
\end{equation}
The first term on the right hand side is the angular momentum change due to mass accretion. After being multiplied by $R^{2}$, it is defined as
\begin{equation}
AM_{\dot{M}}(R)=\dot{M}R \partial_{R} (Rv_{K})\,,
\end{equation}
where $\dot{M}=-\langle \Sigma R v_{R}\rangle_{\phi}$. The second term on the right hand side is the Reynolds stress gradient
or the angular momentum flux gradient. With the additional factor of $R^2$, it is defined as
\begin{equation}
AM_{FH}=-R\partial_{R}(R^{2}\langle \Sigma v_{R}\delta v_{\phi}\rangle _{\phi})\,.
\end{equation}
The last term is the torque by all forces. With the additional factor of $R^2$, it is defined as
\begin{equation}
T(R)=R^{2} \langle \mathbf{R}\times \mathbf{F}\rangle _{\phi}\,.
\end{equation}
For steady linear waves that are generated by the torque and propagate in disks, the $AM_{\dot{M}}$ term is zero since
linear perturbation cannot change the background state and $\dot{M}$ is zero.
In this case, all angular momentum generated by the torque (the $T(R)$ term) is carried away by the wave
(the $AM_{FH}$ term).
However, when the waves steepen into shocks, part of angular momentum flux carried by the wave
is lost into the background disk so that the disk accretes. Thus, shock dissipation can be quantified by
the difference between the $T(R)$ term and the $AM_{FH}$ term. 
 When the disk is in a quasi-steady state as in our cases,  
the terms on the right hand side of Equation \ref{eq:ang} need to balance each other, so that
\begin{equation}
-AM_{\dot{M}}(R)\approx AM_{FH}(R)+T(R)\,.
\end{equation}
This equation suggests that accretion is not led by the torque alone but by the dissipation $(AM_{FH}(R)+T(R))$
 that is defined as the difference between
the total torque and the angular momentum flux carried away by the wave.

\begin{figure}[ht!]
\centering
\includegraphics[trim=0cm 0cm 0cm 0cm, width=0.5\textwidth]{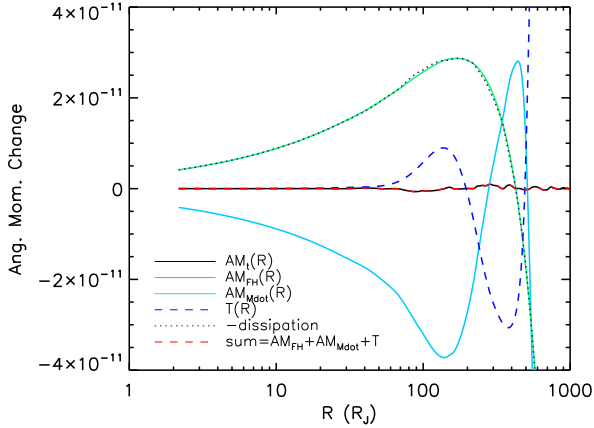} 
\caption{ Angular momentum budget in case Gp2. All quantities are averaged over
every timestep during 50 $\Omega_{5AU}^{-1}<$t$<$ 100 $\Omega_{5AU}^{-1}$. The dotted purple
curve shows the dissipation term $AM_{FH}(R)+T(R)$. The good match between the dissipation
term and the mass accretion term ($AM_{\dot{M}}$) suggests that the shock dissipation indeed drives mass accretion. }
\vspace{-0.1 cm} \label{fig:angbudget}
\end{figure}

The detailed angular momentum budget for case Gp2 is shown in Figure \ref{fig:angbudget}, where all quantities are
averaged over each time step for  50 $\Omega_{5AU}^{-1}$.  The dissipation term
$AM_{FH}(R)+T(R)$ is balanced by the disk accretion $AM_{\dot{M}}(R)$. Figure \ref{fig:angbudget} also shows that most of the torque
exerts on the disk  at $R>50$ R$_{J}$, and part of the torque is used to launch spiral waves. These spiral
arms can propagate all the way to Jupiter surface and  dissipate throughout the whole disk with a non-zero dissipation term. 

When the disk is in a quasi-steady state (as shown in Figure \ref{fig:radial}), we can directly integrate Equation \ref{eq:ang} to derive the relationship between the
mass accretion rate and the stress
\begin{equation}
\dot{M} = \frac{R^2\langle \Sigma v_{R}\delta v_{\phi}\rangle _{\phi}+C-\int R\langle \mathbf{R}\times \mathbf{F}\rangle _{\phi}dR}{R v_{K}}\,, \label{eq:momconst}
\end{equation}
where $C$ is determined by the boundary condition.
Motivated by the standard $\alpha$ disk theory where $\dot{M}=\alpha 3\pi \Sigma c_{s} H$, we divide $3\pi \Sigma c_{s} H$
in the above equation to write
\begin{equation}
\alpha_{eff}=\alpha_{Rey}+\alpha_{const}+\alpha_{T}\,,\label{eq:eff}
\end{equation}
where the effective $\alpha$ calculated by $\dot{M}$ is
\begin{equation}
\alpha_{eff}=\frac{\dot{M}}{3\pi \Sigma c_{s} H}\,,
\end{equation}
the pressure normalized Reynolds stress is
\begin{equation}
\alpha_{Rey}=\frac{\langle \Sigma v_{R}\delta v_{\phi}\rangle _{\phi}}{3\pi \Sigma c_{s}^2}\,,
\end{equation}
and the $\alpha$ due to the boundary condition and the torque are
\begin{align}
\alpha_{const}&=\frac{C}{3\pi \Sigma c_{s} H R v_{K}}\,,\\
\alpha_{T}&=\frac{-\int R\langle \mathbf{R}\times \mathbf{F}\rangle _{\phi}dR}{3\pi \Sigma c_{s} H R v_{K}}\,.
\end{align}

\begin{figure*}[ht!]
\centering
\includegraphics[trim=0cm 0cm 0cm 0cm, width=0.9\textwidth]{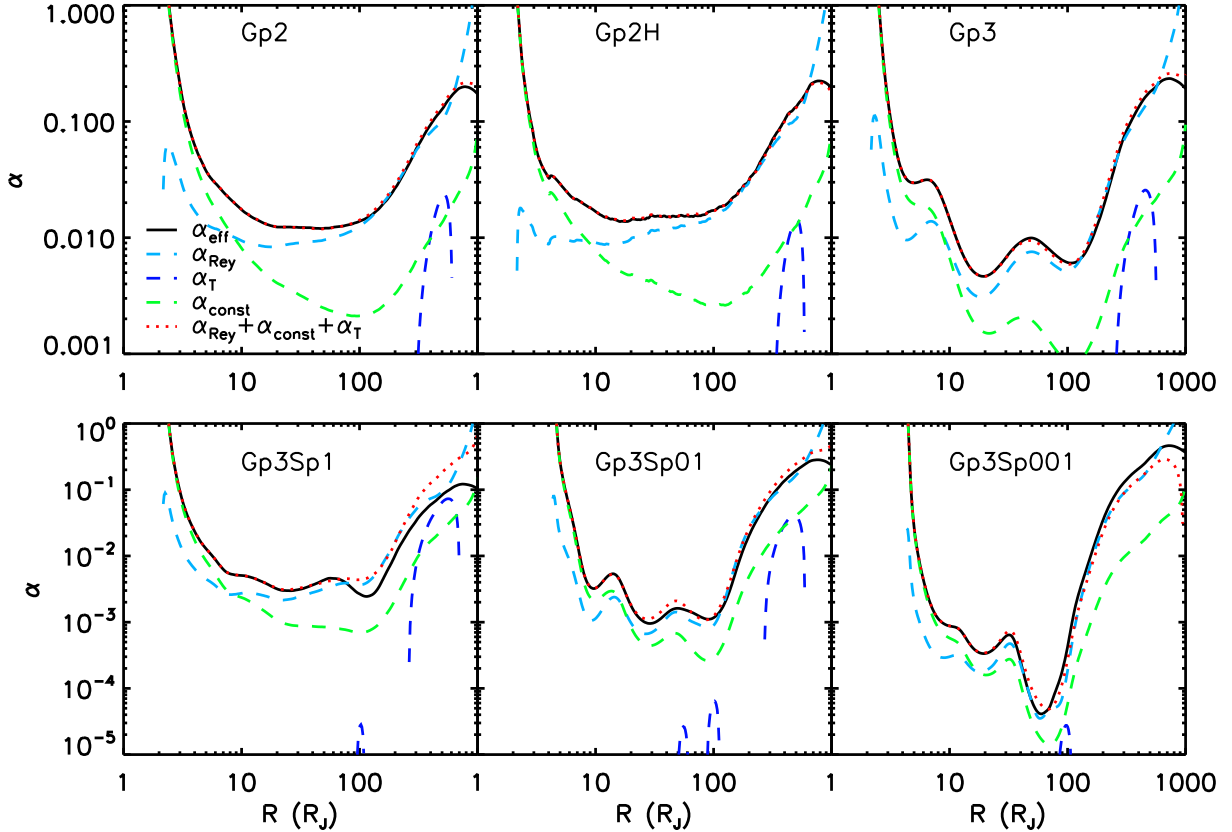} 
\caption{ The time and azimuthally averaged $\alpha$ values with respect to the radius for all the cases.  Generally,
 $\alpha_{eff}$ is at 0.001-0.01 level despite disk accretion rates that span  4 orders of magnitude among different cases. }
\vspace{-0.1 cm} \label{fig:alphaprofile}
\end{figure*}

\begin{figure}[ht!]
\centering
\includegraphics[trim=0cm 0cm 0cm 0cm, width=0.9\textwidth]{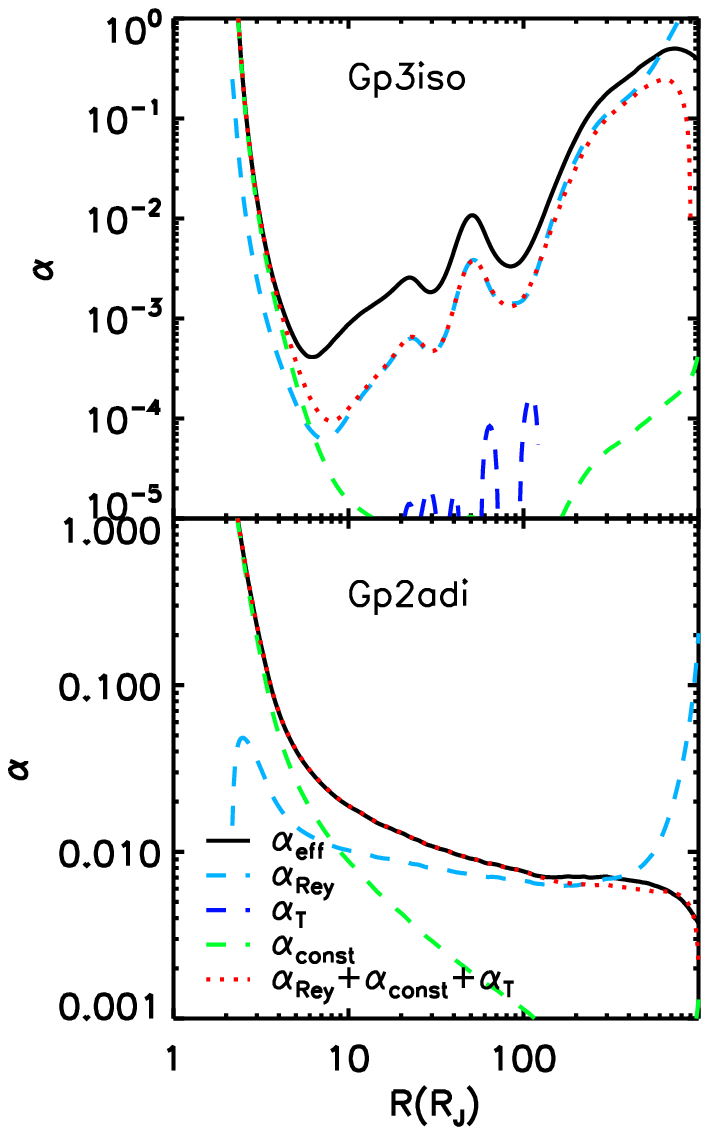} 
\caption{ Similar to Figure \ref{fig:alphaprofile} but for the isothermal and adiabatic cases.}
\vspace{-0.1 cm} \label{fig:alphaprofileiso}
\end{figure}

These $\alpha$ values with respect to the radius are shown in Figure \ref{fig:alphaprofile} for all the cases that have radiative cooling.
{  The isothermal and adiabatic cases are shown in Figure \ref{fig:alphaprofileiso}. For the isothermal case, the black and red curves do not overlap, suggesting that the disk has not reached a steady state. The adiabatic case has profiles similar to Gp2. }
To derive these $\alpha$ values, various components in Equation \ref{eq:momconst} are averaged over each time step for 
10 $\Omega_{5AU}^{-1}$ (for the high resolution case, Gp2H), 50 $\Omega_{5AU}^{-1}$ (for cases with 2 R$_{J}$ inner boundary), 
or 100 $\Omega_{5AU}^{-1}$ (for cases with 4 R$_{J}$ inner boundary) before the end of the simulations.

{  All these $\alpha$ values can be directly measured in the simulations except $\alpha_{C}$ which depends on $C$ and the boundary
condition used in the simulation. We notice that $\alpha_{eff}$ is
far larger than $\alpha_{Rey}$ and $\alpha_{T}$ at the inner boundary for all the cases in Figure \ref{fig:alphaprofile}. Thus,
 we can apply the zero stress boundary condition to our simulations. Specifically,
we use $\alpha_{eff}=\alpha_{const}$ (Equation \ref{eq:eff}) at the inner boundary to derive 
 $C=\dot{M} R_{in}v_{K, in}$ in Equation \ref{eq:momconst} where $R_{in}$ and $v_{K,in}$ are the radius and the Keplerian velocity at the inner radius. Then with this $C$ we can calculate $\alpha_{C}$ shown in  Figure \ref{fig:alphaprofile}.

We want to emphasize that the zero stress boundary in our simulations is exactly the same boundary condition used in the thin disk theory. The non-zero $\alpha_{C}$ is the generic properties of
the zero stress boundary condition. The large $\alpha_{C}$ value at the boundary does not mean that the disk accretion there is induced by our adopted numerical boundary condition in the simulation. Even using the thin disk theory,  an analytical model with a large $\alpha_{C}$ 
at the boundary can be constructed as long as $\Sigma$ goes to small values at the boundary. In order to understand the effects of boundary condition, we need to choose different boundary conditions or simulate the boundary layer directly, which will be the focus of a future paper. }

Figure \ref{fig:alphaprofile} shows that $\alpha_{eff}$ is quite close to $\alpha_{Rey}$ in most
part of the disk. The effect of $\alpha_{T}$ only becomes noticeable starting from half of the Hill radius outwards (the
Hill radius is 740 $R_{J}$). $\alpha_{const}$ is important at the inner disk within 10 $R_{J}$. $\alpha_{eff}$ 
is at 0.001-0.01 level despite disk accretion rates that span four orders of magnitude
among different models. 

The $\alpha$ value is slightly higher in a disk with a higher accretion rate.
This is due to the feedback from  shock dissipation to disk accretion.
With a higher infall rate, more material is present within the Roche sphere (or Hill sphere). As shown 
in Figure \ref{fig:radial}, the disk is larger with a higher infall rate. Since spiral waves
are easier to be excited in disk regions closer to the Lindblad resonances, the larger disks have
stronger spiral waves. After these stronger waves dissipate in disks, they heat
up the disk more significantly, and the disk temperature becomes higher. Since spiral arms
are more open in hotter disks (Figure \ref{fig:spiralT}), they propagate further into the inner disk and
leads to stronger accretion there.

{  Quantitatively, $\alpha_{eff}$ is $\sim$ 0.01 in Gp2 which has $\dot{M}\sim 10^{-5}$ $M_{J}/yr$ and $H/R\sim 0.25$, while
$\alpha_{eff}$ is $\sim$ 0.001 in Gp3Sp01 which has $\dot{M}\sim 4\times10^{-8}$ $M_{J}/yr$ and $H/R\sim 0.1$. Larson (1990) 
derived the $\alpha$ value from a steady, self-similar shock (see also Spruit 1987) as
\begin{equation}
\alpha\sim0.013 [(c_{s}/v_{K})^3+0.08(c_{s}/v_{K})^2]^{1/2}
\end{equation}
Using this equation, we can estimate $\alpha$ to be $\sim 0.002$ when $c_{s}/v_{K}=H/R\sim 0.25$, and 
$\alpha$ to be $\sim 0.0006$ when $H/R\sim 0.1$. These estimates are similar to the simulated values
within the same order of magnitude, which
suggest that 
  angular momentum transport by spiral shocks can indeed be efficient in CPDs.
  On the other hand, these estimated values are a factor of 5 and 2 times smaller than those derived in simulations.
Shocks in our simulations are neither steady or self-similar.
To  understand the difference between simulated values and analytical estimates, a more detailed
analytical model needs to be developed.
}

{  
\subsection{Energy Budget}
Besides studying angular momentum transport by spiral shocks, we also investigate how energy is dissipated
and transported in CPDs. This energy budget analysis is more tedious than the angular momentum budget analysis,
so that we leave it to the Appendix. Overall, since the disk is relatively thick with a large H/R, the radial advection of energy becomes important and the disk generates less infrared 
radiation than that from the thin disk theory by a factor of $\sim$2.}

\section{Discussion}

\subsection{Observational Signatures}
With the CPD structure uniquely determined in our simulations,
we can calculate the observables for CPDs and test if current telescopes
can detect CPDs.

\subsubsection{SEDs at Optical and Near-IR}
\begin{figure}[ht!]
\centering
\includegraphics[trim=0cm 0cm 0cm 0cm, width=0.5\textwidth]{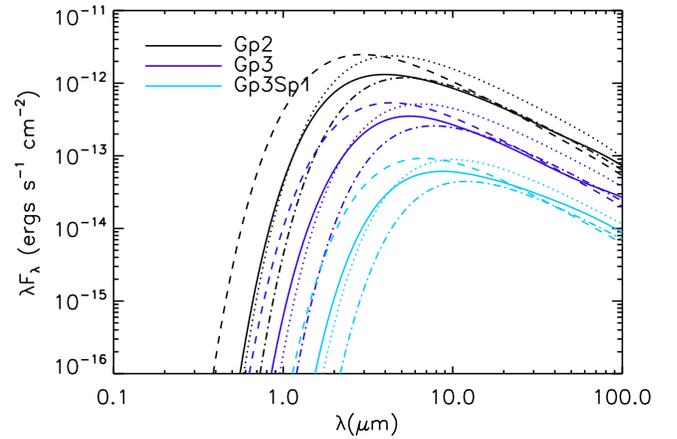} 
\caption{ SEDs generated using the effective temperature from simulations (solid curves, averaged over 50 $\Omega_{5 AU}^{-1}$) 
compared with various analytical models assuming the source is 100 pc away from us (dotted curves: the thin disk approximation; 
dash-dotted curves: the thin disk approximation but only half of the disk accretion rates;
dashed curves:  using the empirical temperature relationship given at Equation \ref{eq:teff}).}
\vspace{-0.1 cm} \label{fig:SED}
\end{figure}

Detailed SEDs of accreting CPDs have been calculated by Zhu (2015) when the disk extends to the planet surface
or when the disk is truncated by magnetic fields of the young planets.
These SEDs are much redder than SEDs produced by brown dwarfs or planets (e.g. Spiegel 
\& Burrows 2012) since the outer disk emits significant infrared flux.
Thus, to use direct imaging to find  the accretion disks around low mass planets and distinguish them
from brown dwarfs or hot high mass planets, it is crucial  to obtain photometry at mid-infrared bands ($L'$, $M$, $N$ bands).

Direct imaging observations have found several red sources within circumplanetary disks (Kraus 
\& Ireland 2012, Quanz \etal 2013, Biller \etal 2014, Reggiani \etal 2014, Currie \etal 2015). Due to their redness,
these sources could be potential candidates for CPDs. Thus, we have used the theoretical SEDs in Zhu (2015)
to fit the observed photometry, trying to constrain the disk properties. 
Table \ref{tb:3} summarizes recent detections of point sources in protoplanetary disks and the predictions from theoretical models. 
We fit the L band observations using the L band magnitudes given in Zhu (2015) to constrain the $M\dot{M}$.
Then we give predicted magnitudes at other wavelength bands. The distance and disk inclination
are from the given references for each source.
To simplify the fitting procedure, we only use the SED models with the disk inner radius of 1 Jupiter radius. 
Even with this simplification, the agreement between the model and observations at various bands is quite good, especially
considering that L band observation is the only one used in the model to make predictions  at all the other bands. 
If the assumption of the inner disk radius is relaxed, 
multiple bands data can be fitted much better 
(e.g. as done in Currie \etal 2015 for HD 100546b.  Extinction also plays a role at H band for  HD 100546b). 
Generally, for all these CPD candidates, $M\dot{M}$ is around 10$^{-6}$ to 10$^{-5}$ M$_{J}^{2}/yr$.

\begin{table*}
\begin{center}
\caption{Magnitudes of Accreting Circumplanetary Disks compared with Observations \label{tab1}}
\begin{tabular}{cccccccccc}
\hline\hline
&$M\dot{M}(M_{J}^2/yr)$ &  J  & H & K & L' & M & N  & 870 $\mu$m & 1.3mm\\ 
\hline
\multicolumn{1}{l}{HD169142b}\\
\hline
Obs.$^{a}$ & &$>$13.8 &   &   & 12.2$\pm$0.5 &  &  & & \\
The. & 10$^{-5}$ &14.8 & 14.66 & 13.82 & 12.2 & 11.62 & 10.12  & $\sim$300 $\mu$Jy & $\sim$100 $\mu$Jy \\
\hline
\multicolumn{1}{l}{HD100546b }\\
\hline
Obs.$^{b}$ & & &19.4$\pm$0.32 & $>$15.43$\pm$0.06 & 13.92$\pm$0.1 & 13.33$\pm$0.16 &  & &\\
The. & 2$\times$10$^{-6}$ & 20.66 & 18.41 & 16.50 & 13.9 & 13.05 & 11.37  & $\sim$800 $\mu$Jy  & $\sim$300 $\mu$Jy\\
\hline
\multicolumn{1}{l}{LkCa 15b}\\
\hline
Obs.$^{c}$ & &  &   & 14.2$\pm$0.5 & 13.2$\pm0.5$ &  &   & & \\
The. & 7$\times$10$^{-6}$ &16.09 & 15.92 & 14.96 & 13.2 & 12.54 & 11.04 & $\sim$300 $\mu$Jy & $\sim$100 $\mu$Jy \\
\hline
\end{tabular}
\label{tb:3}
\end{center}
\tablenotetext{1}{Reggiani \etal 2014, distance of 145 pc, inclination of 0$^{o}$}
\tablenotetext{2}{ Quanz \etal 2015,Currie \etal 2015, distance of 100 pc, inclination of 40$^{o}$}
\tablenotetext{3}{ Sallum \etal 2015, distance of 145 pc, inclination of 50$^{o}$}
\end{table*}

To derive SEDs for CPDs, Zhu (2015) has used the
 temperature profile  from the thin disk theory as in Equation \ref{eq:alphateff} 
where $R_{in}$ is the inner radius of the disk (it equals to $R_{p}$ if the disk extends to the planet surface
or equals to the magnetic truncation radius when the planet has strong magnetic fields and is undergoing magnetospheric accretion, Lovelace \etal 2011.)
\footnote{
Equation \ref{eq:alphateff} was modified in Zhu (2015) in a way that, when the radius is smaller than 1.36 $R_{in}$, 
the temperature is constant and equal to $T = T_{max}$ to heuristically approximate the boundary layer. This
approximation could fit the SED from the accretion disk of FU Ori reasonabley well (Zhu \etal 2007).} .

However, as shown in Appendix, the cooling rate from the thin disk approximation (Equation \ref{eq:alphateff}) is inadequate
to reproduce the real cooling rate from the simulation  due to the radial energy advection. 
To understand how this difference affects the disk's SED, we generate the SEDs directly from simulations. 
First, we use the 
cooling rate in simulations to calculate the disk effective temperature at each radius. Then we assume
blackbody emission at that radius to calculate the SED generated by that annulus of the disk. Finally, we integrate all the disk
emission from the inner boundary to the planet's Hill radius to get the SED of the entire CPD.
These SEDs are plotted in Figure \ref{fig:SED} as solid curves. For comparison, two SEDs calculated
from the thin disk approximation (Equation \ref{eq:alphateff}) are also plotted:
the dotted curves use the same mass accretion rates as those measured from simulations, while the dash-dotted curves assume
only half of the disk accretion rates. Clearly, the dotted curves roughly fit the simulated SEDs at 
the short wavelengths (below where the flux peaks). The dash-dotted curves fit the simulated SEDs
at the long wavelengths (the Rayleigh-Jeans part of the SEDs). This is expected from 
Figure \ref{fig:coolingratio} where at the outer disk the emission is only half of that from the thin disk theory. 
Thus, if we use the Rayleigh-Jeans part of the observed SED to estimate the disk accretion rate based on the thin disk theory, we will 
underestimate the disk accretion rate by a factor of 2. Thus, the mass accretion rates given in Table \ref{tb:3} 
need to be multiplied by a factor of 2 to derive the real disk accretion rates.

The dashed curves in Figure \ref{fig:SED} are SEDs calculated using 
our empirical temperature relationship (Equation \ref{eq:teff}).  This empirical relationship 
works well for the SED at long wavelengths but it over predicts the flux at short wavelengths,
due to the reason presented at the end of Appendix. 

Note that SEDs in Figure \ref{fig:SED} are only from the accretion disks. 
The planet's SEDs have not been added. If the planet has a cold atmosphere as in the ``cold-start'' model (Marley \etal 2007),
its flux is significantly lower than the flux from the disk (Zhu 2015).
Since the zero stress boundary condition has been used to derive the temperature in the thin disk theory
and this boundary condition also stands in our simulations, it implicitly assumes that
the planet rotates at the breakup speed and the disk joins the planet smoothly. 
In reality, the planet may rotate slower than the disk and a boundary layer is formed around the planet,
or the planet has a strong magnetic field to truncate the CPD.
The observational signatures of the boundary layer and magnetospheric accretion will be discussed in \S 4.1.3.

\subsubsection{Submm/mm Flux at ALMA/EVLA Bands}

\begin{table}[ht]
\begin{center}
\caption{Disk Properties and Derived Submm/mm Flux \label{tab1}}
\begin{tabular}{cccccccc}
\tableline
\tableline
\tableline
Run    &  $\dot{M}$ & CPD Mass & Flux$^{a}$ 870$\mu$m   & 1.3 mm & 7mm \\
           &    M$_{J}/yr$   & M$_{J}$ &$\mu$Jy & $\mu$Jy & $\mu$Jy \\
\tableline
Gp2  & 1.07$\times$10$^{-5}$ & 6.4$\times$10$^{-4}$ & 200 & 79 & 1.2 \\
Gp3 & 2.26$\times$10$^{-6}$  & 3.1$\times$10$^{-4}$ & 84 & 34 & 0.51 \\
Gp3Sp1 & 3.89$\times$10$^{-7}$ & 1.0$\times$10$^{-4}$ & 31 & 12 & 0.12\\
Gp3Sp01 & 3.93$\times$10$^{-8}$  & 3.6$\times$10$^{-5}$ & 12 & 4.0 & 0.028\\
Gp3Sp001 & 1.92$\times$10$^{-9}$  & 2.0$\times$10$^{-5}$ & 6 & 2.1 & 0.013\\
Gp3iso & -  & 1.0$\times$10$^{-3}$ & 44 & 16 &  0.28\\
\tableline
\end{tabular}\label{tb:2}
\end{center}
\tablenotetext{1}{ The total received flux is calculated by assuming the source is at 100 pc.}
\end{table}

With ALMA and VLA's great sensitivity, we may detect CPDs at submm/mm bands. 
Isella \etal (2014) have used parameterized models to calculate submm/mm flux for CPDs and 
suggest that ALMA can probe CPDs with mass down to 5$\times$10$^{-4}$ M$_{J}$. Due to the use
of the parameterized models, disk mass, size, and accretion rate are degenerate, leaving a large parameter
space to explore. In our  first-principle calculations, the disk density and temperature structure are uniquely determined at a given
accretion rate,
so that we can give a unique prediction of the CPDs' submm/mm flux.

To calculate the disk's flux at submm/mm, we use the dust opacity of 
0.034$\times (0.87 mm/\lambda)$ cm$^{2}$/g (Andrews \etal 2012). At one annulus, if the disk
is optically thin at the given wavelength, we approximate the brightness temperature ($T_{b}$) as 
the product of the midplane temperature 
and the optical depth. When the disk is optically thick at the given wavelength,
the brightness temperature is chosen as the temperature where the disk becomes optically thick
at that wavelength.
In detail, the temperature at the optical depth of $\tau_{R}$ is
\begin{equation}
T(\tau_{R})^4=\frac{3}{16}T_{eff}^4 \tau_{R}+T_{\rm ext}^4\,,\label{eq:Ttau}
\end{equation}
where $ \tau_{R}$ is the optical depth calculated using the Rosseland mean opacity.
At the disk height where $\tau_{R}=\kappa_{R}/\kappa_{\lambda}$, the disk becomes optically thick at $\lambda$.
If we plug the effective temperature from Equation \ref{eq:cooling} into Equation \ref{eq:Ttau}, we can derive
\begin{equation}
T_{b}^4=T(\tau_{R}=\kappa_{R}/\kappa_{\lambda})^4=(T_{c}^4-T_{\rm ext}^4)/\tau_{\lambda}+T_{\rm ext}^4\,.
\end{equation}
where $\tau_{\lambda}=\Sigma \kappa_{\lambda}/2$ \footnote{Since $\kappa_{\lambda}$ is normally much smaller than 
$\kappa_{R}$ at submm, we have assumed $\Sigma \kappa_{R}\gg1$ to simplify the derivation above.}.
Then knowing $T_{b}$ at each radius, we can integrate the emission from the whole disk to derive the total flux.
The flux at various wavelengths for disks with different accretion rates is given in Table  \ref{tb:2}. The source
is assumed to be 100 pc away from us. 

Although the CPDs in our simulations are 5 AU away from their central stars, we can roughly scale our
results to CPDs at other distances. Based on Figure \ref{fig:radial}, the disk surface density is almost a constant along radii,
and the midplane temperature roughly scales as $R^{-1}$. Thus the total intensity scale as $R$. Since the size
of the CPD is 1/3 of the planet's Hill radius (Martin \& Lubow 2011a), the total submm/mm flux in Table \ref{tb:2} should roughly
scale as the distance of the planet to the central star. 

Thus, we scale the derived 
submm/mm flux for Jupiter at 5 AU in Table \ref{tb:2} to several 
CPD candidates and provide their fluxes in Table \ref{tb:3}.
We want to caution that this scaling is highly simplified. Intrinsically, it assumes that $\alpha$ in the CPD around Jupiter is the same 
as CPDs around young planets at 20-50 AU. Considering the complicated interplay between the thermodynamics, spiral excitation,
and shock dissipation, this is unlikely to be true. Furthermore, we implicitly assume that the CPD candidates all have 1 Jupiter mass planet
at the center, the same as in our simulations. Thus, the submm flux given in Table \ref{tb:3} can only be considered as an order of
magnitude estimate. Nevertheless, with ALMA's great sensitivity (assuming 1.4 mm and all bandwidth set for continuum, we can detect 230 $\mu$Jy point source with the signal-to-noise ratio of 10 in 57 minutes, private communication with John Tobin), we should easily detect these sources
if their origins are CPDs.

\subsubsection{Other Observables and Variability}
Our simulations suggest that CPDs can accrete quite efficiently due to the spiral shocks. Besides the SED,
optical/near-IR emission lines (e.g. Hartmann \etal 1994, Muzerolle \etal 1998, 2001), UV excess 
(e.g. Gullbring \etal 2000, Herczeg \& Hillenbrand 2008, Ingleby \etal 2013),  and line veiling (e.g. Calvet \& Gullbring 1998) 
are all accretion signatures (Bouvier \etal 2007, Rigliaco \etal 2012, Alcal{\'a} \etal 2014). For disks around young stellar objects, the profiles of emission lines 
suggest that these lines are produced by infalling material during magnetospheric accretion (Hartmann \etal 1994). UV excess
is also believed to be produced by the accretion hot spots on the surface of the star where the magnetosphere connects to the star (Calvet \& Gullbring 1998). 
If CPDs around young planets can undergo magnetospheric accretion, they should also produce these accretion features. However, in order to 
truncate accreting CPDs by magnetic fields, young planets need to have magnetic fields $\sim$ 10-100 gauss 
(Fendt 2003, Lovelace \etal 2011, Zhu 2015),  much stronger than Jupiter's current magnetic field. If young planets do not have
such strong magnetic fields, the disk will accrete onto the planet through the boundary layer. Although accretion through 
the boundary layer has been better understood (Belyaev et al. 2013, Philippov \etal 2016), boundary layers in CPDs can be 
quite thick ($H/R\sim$0.2, Figure \ref{fig:radial}) and directly exchange energy with the planet (Owen \&Menou 2016). 
Whether boundary layers can
still produce strong emission lines and UV excess need to be explored. 

\begin{figure}[ht!]
\centering
\includegraphics[trim=0cm 0cm 0cm 0cm, width=0.5\textwidth]{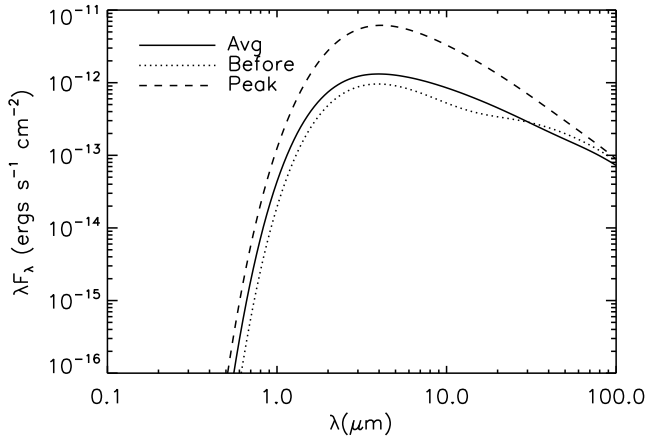} 
\caption{ Similar to Figure \ref{fig:SED}, the solid curve is the SED generated using the effective temperature which has
been averaged over 50 $\Omega_{5 AU}^{-1}$. The dotted curve is SED from the snapshot before the outburst and the 
dashed curve is the SED during the peak of the outburst. These two snapshots are only separated by 1.8 years.}
\vspace{-0.1 cm} \label{fig:SEDvariable}
\end{figure}

Our simulations also suggest that the disk accretion rate varies significantly at the timescale of years 
when the disk accretion rate is $\sim 10^{-5}M_{J}/yr$. The disk cooling timescale is also around one year for such a disk, estimated
with Equation (8) in Zhu \etal (2015).  Thus the change of the disk accretion rate would lead to the variability of 
the observables over  years timescale.
Thus, for these disks, the accretion tracers
may vary between observations at different epochs. {  This is shown in Figure \ref{fig:SEDvariable} where the dotted
and dashed curves are only separated by 1.8 years but the peak intensity is differed by almost a factor of 10.}

Besides disk accretion, CPDs could also launch
jets/outflows (Quillen \& Trilling 1998, Gressel \etal 2013), and have shock fronts due the inflow from the circumstellar disks (Tanigawa \etal 2012).
They may also have observational signatures.  Last but not least, the Keplerian rotation of the CPD may even be probed by molecular lines using ALMA (Perez \etal 2015).

\subsection{Satellite Formation}
\begin{figure}[ht!]
\centering
\includegraphics[trim=0cm 0cm 0cm 0cm, width=0.5\textwidth]{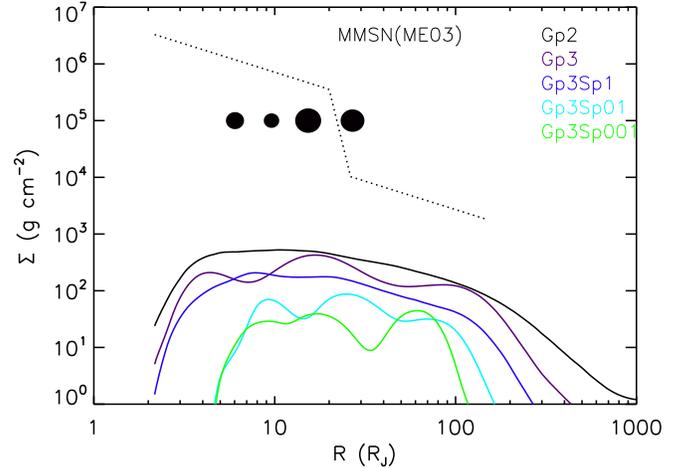} 
\caption{ The disk surface density from various simulations compared with the "minimum mass sub-nebula'' model in  \cite{mosqueira2003} (dotted curve).
The four dots label the position of Jupiter's four moons: Io, Europa, Ganymede, and Callisto.}
\vspace{-0.1 cm} \label{fig:subnm}
\end{figure}

We can compare the disk surface density from our simulations with the disk surface density required 
by various satellite formation models. Our derived disk surface density is orders of magnitude smaller than
the in-situ satellite formation models, or so-called the ``minimum mass sub-nebula" model (Figure \ref{fig:subnm}). 
Our simulated CPDs have surface densities between 10 and 1000 g cm$^{-2}$ depending on the disk accretion rates, while
the ``minimum mass sub-nebula" \citep{lunine1982, coradini1984, mosqueira2003} has a surface 
density of few$\times 10^{5}$ g cm$^{-2}$. The model in \cite{mosqueira2003} is plotted as the dotted curve
in Figure \ref{fig:subnm}, which is much higher than the surface density from our 2-D simulations. 
On the other hand, the surface densities in our simulations are consistent with that in 
the ``gas-starved'' satellite formation model 
\citep{canup2002,canup2006,ward2010}.

Since Galilean satellites are ice-rich, they need to form in conditions where the disk midplane temperature
is below ~150 to 200 K, depending on pressure (Prinn \& Fegley 1989, Canup \&Ward 2002). In order for the
disk to be cooler than 200 K at R$>30 R_{J}$, Figure \ref{fig:radial} suggests that the disk accretion rate
 needs to be smaller than $4\times 10^{-8} M_{J}/yr$ or $10^{-5} M_{\oplus}/yr$.
Thus, Galilean satellites should form in a CPD with a very low accretion rate.  
This accretion rate is broadly consistent with that (2$\times$10$^{-7}M_{J}/yr$) suggested by Canup \& Ward (2002).
Using $\alpha\sim$0.001, Heller \& Pudritz (2015ab) have reached the similar conclusion that Jovian moons 
have to form during the final stages of CPD accretion.   Despite this low accretion rate,
it will only take $\sim 5\times10^{5}$ yr, well within the protoplanetary disk lifetime, 
for such a CPD to supply enough material for building Galilean satellites.

\begin{figure}[ht!]
\centering
\includegraphics[trim=0cm 0cm 0cm 0cm, width=0.5\textwidth]{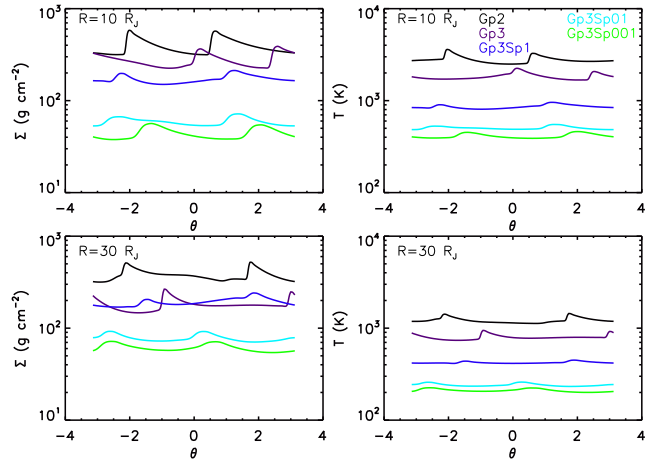} 
\caption{The 1-D density (left panels) and temperature (right panels)
profiles along the azimuthal direction at 10 (upper panels) and 30 (lower panels) $R_{J}$ for various cases. }
\vspace{-0.1 cm} \label{fig:aziprofile}
\end{figure}

Although the effects of spiral shocks can be 
represented by the equivalent $\alpha$ parameter as shown above, spiral shocks are substantially different from viscosity in 
that they can travel long distance in disks and have non-homogeneous  azimuthal structure. 
The density and temperature are high at the shock
front. Since the spiral shocks are stationary in the corotating frame, material in the CPD will move in and out of the spiral
shocks during each orbit around the planet. Disk material can frequently be heated up, dissociated, and condensed again.
We plot the 1-D density and temperature 
profiles along the azimuthal direction at 10 and 30 $R_{J}$ in Figure \ref{fig:aziprofile}.
Thus, disk material is heated episodically (twice every orbit due to the two spiral arms) and may have
imprints in the material accreted onto the satellites or left in the CPD. 

\subsection{Caveats}
MHD effects have been completely ignored in this study.  As shown in Figure \ref{fig:radial}, for disks with accretion rate larger
than $\sim$10$^{-7}M_{J}/yr$, the inner disk within 10-50 $R_{J}$ can be hot enough ($>1000$ K)  to 
sustain MRI (Keith \& Wardle 2014).  On the other hand, 
$\alpha$ due to MRI is similar to the $\alpha$ by the spiral shocks ($\alpha\sim$ 0.001 - 0.1). Thus,
the operation of MRI at the inner disk may not dramatically change the disk structure. 

Previous studies (Fujii \etal 2011, 2014, Keith \& Wardle 2014)
find that it is difficult to sustain MRI in the majority part of the CPD beyond 30 $R_{J}$.
Non-ideal MHD effects coupled with magnetic breaking may be important there. 
Without efficient accretion onto the planet,  mass being supplied from the circumstellar
 disk will pile up in CPDs, leading to gravitational instability
and triggering accretion outbursts (Lubow \& Martin 2012), similar to the outburst mechanism proposed 
for FU Orionis systems (Armitage \etal 2001, Zhu \etal 2009, Martin \& Lubow 2011b).
Here, we propose spiral shocks as another way to drive accretion. How this new mechanism interplays
with previous proposed mechanisms needs further studies. 

One of the biggest caveats in this study is that we are limited to 2-D. Previous studies (Bate \etal 2003, Machida \etal 2008, Tanigawa \etal 2012, 
Morbidelli \etal 2014, Szul{\'a}gyi \etal 2014) have
suggested that CPDs have complicated 3-D structure. Material almost falls vertically from the circumstellar disk
 to the CPD. {   Even in our 2-D simulations, $H/R$ is approaching 0.5 at the planet's Hill radius (Figure \ref{fig:radial}) and
 our thin disk approximation breaks down there. Thus, our simulations may not accurately capture dynamics there. 
 On the other hand,
CPDs are truncated at 1/3-1/2 of the planet's Hill radius and material around the 
planet's Hill radius should not affect dynamics within CPDs.
 Szul{\'a}gyi et al. (2016) has shown that the inner disk can undergo thermal runaway and can reach peak temperature of 10,000 K. In this case,
the whole region within the Hill sphere can even become a hydrostatic 
 envelop. Among our simulations, the case with the highest accretion rate (Gp2) may also undergo thermal runaway if we
 use the realistic molecular opacity and then H/R will be 0.5 at the inner disk. On the other hand, Szul{\'a}gyi et al. (2016) suggests
 that such thermal runaway could be due to the lack of proper treatment for thermal dynamics. Nevertheless, for our 
 other cases except Gp2, $H/R$ is less than 0.25 in most regions of the disk and the thin disk approximation may still be valid.
 Spiral shocks have complicated 3-D structure (Zhu \etal 2015, Lyra \etal 2016) and can even be unstable in 3-D (Bae \etal 2016).
A proper study on how shock driven accretion is affected by the 3-D effects needs future
3-D simulations with realistic radiative transfer and thermodynamics.
}

\section{Conclusion}
 Circumplanetary disks (CPD) control the growth of  planets, supply material for satellites to form, and
provide observational signatures of young forming planets. In this paper, we provide a new
mechanism to explain how CPDs accrete. 

We have carried out two dimensional hydrodynamical
simulations to study CPDs using Athena++. Simple radiative cooling has been considered in the simulation. 
Different from most previous simulations, we choose a coordinate system centered on the planet, which
significantly reduces the numerical error and enables simulations with small numerical viscosity. 
Our simulation domain extends from the circumstellar disk
all the way to the Jupiter's surface. A gap in the circumstellar disk is prescribed, allowing us to control
the inflow rate from the circumstellar to the circumplanetary disk.  

Two spiral shocks are present in CPDs, induced by the tidal force from the central star.
We find that these spiral shocks can lead to significant angular momentum transport and energy dissipation in CPDs. 
 {  Meanwhile, dissipation and heating by spiral shocks have a positive feedback on shock-driven accretion itself.}
 As the disk is heated up by spiral shocks, the shocks become more open and propagate further into the inner disk, leading
to more efficient angular momentum transport at the inner disk. On the other hand,
shock driven accretion cannot guarantee strict steady disk accretion since the angular momentum transport
depends on the global wave propagation and local shock dissipation. Mass will pile up in some regions of the disk and
sometimes vortices are produced, which produce short-timescale (months to years) variability. The disk is adjusting itself through
these variabilities and tries to maintain a quasi-steady state. 
Eventually, a quasi-steady state of accretion flow is reached in our simulations 
from the planet's Hill radius all the way to the planet surface. After averaging quantities over a long timescale,
angular momentum budget is carefully analyzed. 
The effective $\alpha$-coefficient characterizing angular momentum transport due to spiral shocks is $\sim$0.001-0.02
even though the disk accretion rates span  4 orders of magnitude. {  The $\alpha$ value is higher in a disk 
with a higher accretion rate due to the shock heating feedback.} 

With energy budget analysis,
we show that radial advection of energy becomes important and the disk generates less infrared 
radiation than that from the thin disk approximation by a factor of $\sim$2. 
 Thus, if we use the infrared flux calculated from the thin
disk approximation to derive the accretion rate of CPDs, we may underestimate the disk accretion rate by a factor of 2.

Finally, we calculate
the flux from CPDs at ALMA and EVLA wavelength bands and predict the flux to several recent CPD candidates (e.g. HD169142b,
HD100546b, LkCa 15b). These CPD candidates should be relatively bright at ALMA wavelength bands.
In future, ALMA should be able to discover many accreting CPDs.

Unlike near-IR emission which comes from the inner disk region,
submm flux comes from the outer disk region which is less subject to vortex production and destruction.
Furthermore the dynamical time is much longer at the outer disk. Thus 
submm flux will be 
more steady over time than the optical/near-IR flux. We may see that CPDs appear and disappear 
between two epoch optical/near-IR observations while they are  bright all the time at ALMA bands.

Although our simulations are limited to 2-D, the possibility that we may have understood how CPDs accrete
and its huge implications on observations and satellite formation make it worth being studied in detail in future.

\acknowledgments
We thank useful comments from Lee Hartmann, and Nurial Calvet. 
We thank the referee for a helpful and constructive report. 
All hydrodynamical simulations are carried out using computer supported by the 
Princeton Institute of Computational Science and Engineering, and the Texas Advanced Computing Center (TACC) 
at The University of Texas at Austin through XSEDE grant TG-
AST130002.
Part of the work is done when Z.Z. is at Princeton, supported by
NASA through Hubble Fellowship grant HST-HF-51333.01-A
awarded by the Space Telescope Science Institute, which is
operated by the Association of Universities for Research in Astronomy, Inc., for NASA, under contract NAS 5-26555.

\appendix

\section{Energy Budget}
Besides studying angular momentum transport by spiral shocks, we also investigate how energy is dissipated
and transported in CPDs.  
Understanding the energy budget during the accretion process is essential for studying the disk thermal
structure and the observational signatures of disks (e.g. Spectral Energy Distributions). CPDs are quite geometrically thick with the aspect ratio reaching 0.3 (Figure \ref{fig:radial}) so that local cooling may not perfectly balance local viscous/shock heating and radial advection of energy can be important. 

\begin{figure}[ht!]
\centering
\includegraphics[trim=0cm 0cm 0cm 0cm, width=0.5\textwidth]{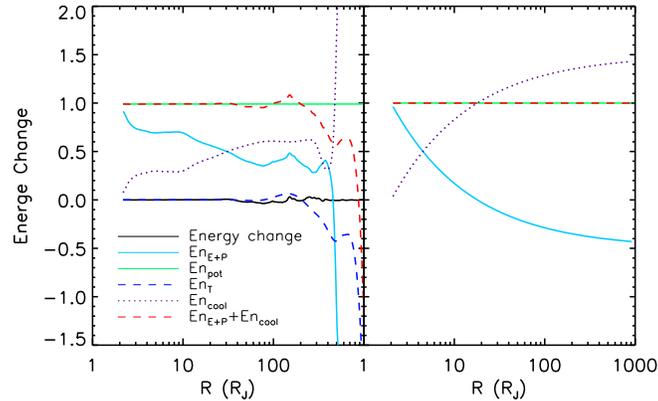} 
\caption{Energy budget terms in case Gp2 (the left panel) and thin disk approximation (the right panel). 
The quantities are averaged over
every timestep during 50 $\Omega_{5AU}^{-1}<$t$<$ 100 $\Omega_{5AU}^{-1}$.   }
\vspace{-0.1 cm} \label{fig:energybudget}
\end{figure}

To calculate the energy budget, we take a similar approach as  for the angular momentum equation by averaging the energy equation in the azimuthal direction
\begin{align}
\partial_{t}\langle E\rangle _{\phi}+\frac{1}{R}\partial_{R}[\langle  R(E+P)v_{R}\rangle _{\phi}]\nonumber\\
=-\langle \Sigma\nabla \Phi_{p}\cdot \mathbf{v}\rangle _{\phi}+\langle \mathbf{F}\cdot \mathbf{v}\rangle _{\phi}-\langle Q_{cool}\rangle_{\phi}\,.\label{eq:energyavg}
\end{align}
The second term on the left hand side is the radial energy flux gradient which is designated as En$_{E+P}$. The first term on the right hand side, which is the gravitational potential energy released during the accretion process,
 is En$_{pot}$. The second term on the right hand side is the work done by the torque, En$_{T}$, while the cooling rate $\langle Q_{cool}\rangle_{\phi}$ is En$_{cool}$. Thus, the above
 equation is
\begin{equation}
\partial_{t}\langle E\rangle _{\phi}+{\rm En}_{E+P}={\rm En}_{pot}+{\rm En}_{T}-{\rm En}_{cool}\,,
\end{equation}
which basically states that the energy change is due to the flux into the domain, the work done by forces (the gravitational force from the central object and other forces), and radiative cooling.
After being normalized with $GM\dot{M}/R^3$, all these terms for case Gp2 are plotted in the left panel of Figure \ref{fig:energybudget}. Figure \ref{fig:energybudget} suggests that the work done by the torque is negligible at the inner disk and the cooling rate equals the difference between the released potential energy and the energy flux gradient.

After removing second order terms, En$_{E+P}$ can be expanded as
\begin{align}
{\rm En}_{E+P}=&\frac{1}{R}\partial_{R}[\langle R(E+P)v_{R}\rangle _{\phi}]\nonumber\\
\sim&\frac{1}{R}\partial_{R}[\langle Rv_{R}(\frac{1}{2}\Sigma v_{K}^2+\Sigma v_{K}\delta v_{\phi}+\frac{\gamma}{\gamma-1}P)\rangle_{\phi}]\nonumber\\
\sim&\frac{1}{R}\partial_{R}[R(\frac{1}{2}v_{K}^2\langle \Sigma v_{R} \rangle_{\phi}+v_{K}\langle\Sigma v_{R}\delta v_{\phi}\rangle_{\phi}+\frac{\gamma}{\gamma-1}\langle v_{R}P \rangle_{\phi})]
\end{align}
With $\dot{M}=-R\langle \Sigma v_{R}\rangle_{\phi}$, the above equation becomes
\begin{equation}
{\rm En}_{E+P}\sim\frac{1}{R}\partial_{R}[-\frac{1}{2}v_{K}^2\dot{M}+R v_{K}\langle\Sigma v_{R}\delta v_{\phi}\rangle_{\phi}+\frac{\gamma}{\gamma-1}R \langle v_{R}P \rangle_{\phi}]\label{eq:energysimple}
\end{equation}
The three terms on the right hand side are the gradient of  kinetic energy, Reynolds stress, and the pressure.
Thus, we designate the three terms as En$_{kic}$, En$_{Rey}$, and En$_{pre}$, and we have
\begin{equation}
{\rm En}_{E+P}\sim{\rm En}_{kic}+{\rm En}_{Rey}+{\rm En}_{pre}\,.
\end{equation}
Note that ${\rm En}_{pre}$  actually consists of two parts: one is the advection of the internal energy and the other is related to the work done by the pressure. 

If we assume that the disk accretion rate is a constant, we can replace the Reynolds stress with the disk accretion rate by using Eq. \ref{eq:momconst} 
\begin{align}
{\rm En}_{Rey}=&\frac{1}{R}\partial_{R}(R v_{K}\langle\Sigma v_{R}\delta v_{\phi}\rangle_{\phi})\nonumber\\
=&\frac{1}{R}\partial_{R}[ v_{K}^2\dot{M}-v_{K}\frac{C}{R}+\frac{v_{K}}{R}\int R\langle \mathbf{R}\times \mathbf{F_{ext}}\rangle _{\phi}dR]\\
=&{\rm En}_{\dot{M}}+{\rm En}_{Const}\nonumber\,,
\end{align}
where we have further neglected the external force term, and designate the first and second term as En$_{\dot{M}}$ and En$_{Const}$.
Thus, finally we have
\begin{equation}
{\rm En}_{E+P}={\rm En}_{kic}+{\rm En}_{\dot{M}}+{\rm En}_{Const}+{\rm En}_{pre}\,.
\end{equation}
If we
assume $\phi_{p}=-v_{K}^2$ and $v_{K}=\sqrt{GM/R}$, and normalize Equation \ref{eq:energyavg} with $GM\dot{M}/R^3$ (designate the corresponding quantities with a hat symbol), we have
\begin{equation}
\overline{{\rm En}_{cool}}+\overline{{\rm En}_{kic}}+\overline{{\rm En}_{\dot{M}}}+\overline{{\rm En}_{Const}}+\overline{{\rm En}_{pre}}=\overline{En_{pot}}+\overline{En_{T}}\nonumber\,,
\end{equation}
and thus
\begin{align}
\overline{<Q_{cool}>_{\phi}}+\frac{1}{2}+(-1+\frac{3}{2}\frac{C}{\dot{M}(GMR)^{1/2}})\nonumber\\
+\frac{R^2}{GM\dot{M}}\partial_{R}(\frac{\gamma}{\gamma-1}R \langle v_{R}P \rangle_{\phi})=1\,.
\end{align}

If we assume zero stress at $R_{in}$ and neglect the pressure term, as in the thin disk approximation, 
we have derived
$C=\dot{M}R_{in}v_{K,in}$ in \S 3.2.
Thus 
\begin{equation}
\overline{<Q_{cool}>_{\phi}}=\frac{3}{2}-\frac{3}{2}(\frac{R_{in}}{R})^{1/2})\,.
\end{equation}
Among the total cooling rate,1 comes from the release of the gravitational energy, and 1/2-3/2$(R_{in}/R)^{1/2}$ comes from the radial energy transport.
Except the very inner disk, the radial energy transport normally heats up the disk. 
-3/2$(R_{in}/R)^{1/2}$ in the radial transported energy is from the zero torque boundary condition (Shakura \& Sunyaev 1973).

To compare all the energy terms derived directly from our simulations with the corresponding terms in the thin disk approximation, 
we summarize all the normalized energy terms in the thin disk theory
\begin{align}
\overline{En_{cool}}=&\overline{En_{pot}}+\overline{En_{T}}-\overline{En_{E+P}}\label{eq:Cool}\\
\overline{En_{E+P}}=&\overline{En_{kic}}+\overline{En_{\dot{M}}}+\overline{En_{Const}}+\overline{En_{pre}}\label{eq:EpP}\\
\overline{En_{pot}}=&1\\
\overline{En_{T}}=&0\\
\overline{En_{kic}}=&\frac{1}{2}\\
\overline{En_{pre}}=&0\\
\overline{En_{\dot{M}}}=&-1\\
\overline{En_{Const}}=&\frac{3}{2}\left(\frac{R_{in}}{R}\right)^{1/2}\,.
\end{align}
Thus, 
\begin{align}
\overline{En_{E+P}}=&-\frac{1}{2}+\frac{3}{2}\left(\frac{R_{in}}{R}\right)^{1/2}\\
\overline{En_{cool}}=&\frac{3}{2}-\frac{3}{2}\left(\frac{R_{in}}{R}\right)^{1/2}\,.
\end{align}
and, eventually,
\begin{equation}
\sigma T_{eff}^4 = {3 G M_{p}\dot{M} \over 8\pi R^{3}}
\left(1-\left(\frac{R_{in}}{R}\right)^{1/2}\right)\,,\label{eq:alphateff}
\end{equation}
The normalized quantities in the thin disk approximation are plotted in the right panel of Figure \ref{fig:energybudget}.

\begin{figure}[ht!]
\centering
\includegraphics[trim=0cm 0cm 0cm 0cm, width=0.5\textwidth]{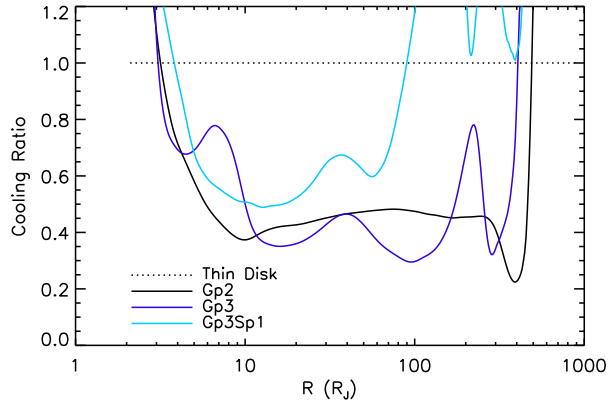} 
\caption{ The ratio between  $\overline{En_{cool}}$ from different simulations and that derived from the thin disk approximation.
Cooling rates are averaged over 50$\Omega_{5AU}^{-1}$. }
\vspace{-0.1 cm} \label{fig:coolingratio}
\end{figure}

By comparing the left and right panels in Figure  \ref{fig:energybudget} where all quantities have been
normalized by 8G$M_{J}$$\dot{M}$/$R^3$, we can see
that the cooling rate in the simulation is smaller than the cooling rate derived from the thin disk
approximation, especially at the outer disk. The ratio between the cooling rate from simulations and that from the thin disk approximation 
is shown in Figure \ref{fig:coolingratio}. Clearly, the cooling rate is only half of that predicted in the thin disk theory at $10 R_{J}<R<100 R_{J}$. 
 Based on the relation in Equation \ref{eq:Cool}, this lower cooling rate in real simulations is caused by 
the higher value of $\overline{En_{E+P}}$. A higher $\overline{En_{E+P}}$ value means that the radial energy transport depletes more energy
from the disk. 

\begin{figure}[ht!]
\centering
\includegraphics[trim=0cm 0cm 0cm 0cm, width=0.5\textwidth]{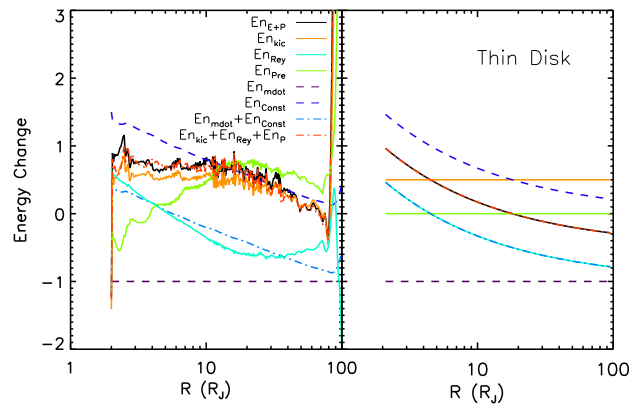} 
\caption{ Different components of $En_{E+P}$ in case Gp2 (the left panel) and the thin disk approximation (the right panel). 
The components are calculated using quantities at t=97$\Omega_{5AU}^{-1}$ when the disk accretion rate is almost a constant throughout the disk.  }
\vspace{-0.1 cm} \label{fig:Enepp}
\end{figure}

To understand why $\overline{En_{E+P}}$ is higher in simulations, we plot different components of  $\overline{En_{E+P}}$
in Figure \ref{fig:Enepp}. By comparing the real simulations (the left panel) with the thin disk approximation (the right panel),
we can see that the pressure term (En$_{pre}$) which represents the radial advection of both the internal energy and the pressure
 is non-negligible in the simulation.   It is negative at the inner disk within 4 $R_{J}$ and positive at the outer disk. Thus,
 the inner disk loses energy and the outer disk gains energy due to the non-zero  En$_{pre}$ term. 
This non-negligible radial transport of energy makes the CPD more like a "slim disk" (Abramowicz et al. 1988) than a thin disk.

We also notice that $\overline{En_{pre}}$ almost balances the Reynolds stress term ($\overline{En_{Rey}}$), so that
the total energy flux gradient term ($\overline{En_{E+P}}$) is almost equal to the kinetic energy term ($\overline{En_{kic}}$).
Thus, empirically, we can approximate $\overline{En_{E+P}}$ as 1/2 instead of $-1/2+3/2\left(R_{in}/R\right)^{1/2}$ in the thin disk theory.
The new cooling rate $\overline{En_{cool}}$ is thus $\sim$1/2. 
If we relate the cooling rate with the effective temperature ($Q_{c}=\sigma T_{eff}^{4}$), we can derive the effective
temperature in the CPD as
\begin{equation}
\sigma T_{eff}^{4}\sim\frac{GM\dot{M}}{8\pi R^{3}}\,,\label{eq:teff}
\end{equation}
which is significantly different from the temperature in the thin disk theory (Equation \ref{eq:alphateff}). 
Thus, compared with the thin disk, the simulated disk is hotter at the inner disk and cooler at the outer disk. 
We want to emphasize that Equation \ref{eq:teff} which assumes $\overline{En_{cool}}=$1/2
is an empirical approximation, it slightly overestimates 
the inner disk temperature compared with real simulations as shown in Figure \ref{fig:energybudget} where $\overline{En_{cool}}$ is smaller than 1/2 at the inner disk.

Overall, our simulations show different temperature profiles from the thin disk theory. This difference is
not caused by the boundary condition since the zero stress boundary condition also stands in our simulations as discussed
at the end of \S 3.2. Instead, the difference is
 caused by
the radial advection of energy since H/R is large and the thin disk approximation is not valid anymore in the CPD.

\end{document}